\pdfoutput=1

\documentclass[aip,jcp,reprint]{revtex4-1}
\usepackage[pdftex]{graphicx}
\usepackage{hyperref}
\usepackage{natbib}
\usepackage{amssymb,amsmath}
\usepackage{amsfonts}
\usepackage{mathrsfs}
\usepackage[english]{babel}
\usepackage{epsfig}
\usepackage{epstopdf}
\usepackage{babelbib}
\usepackage{dcolumn}
\usepackage{bm}

\renewcommand{\Re}{\mathrm{Re}}

\begin{document}

\title{Symmetry effects in electrostatic interactions between two arbitrarily charged spherical shells in the Debye-H\"{u}ckel approximation} 

\author{An\v{z}e \surname{Lo\v{s}dorfer Bo\v{z}i\v{c}}}
\email{anze.bozic@ijs.si}
\affiliation{Department of Theoretical Physics, Jo\v zef Stefan Institute, SI-1000 Ljubljana, Slovenia}
\author{Rudolf \surname{Podgornik}}
\affiliation{Department of Theoretical Physics, Jo\v zef Stefan Institute, SI-1000 Ljubljana, Slovenia}
\affiliation{Department of Physics, Faculty of Mathematics and Physics, SI-1000 Ljubljana, Slovenia}

\date{\today}

\begin{abstract}
Inhomogeneous charge distributions have important repercussions on electrostatic interactions in systems of charged particles but are often difficult to examine theoretically. We investigate how electrostatic interactions are influenced by patchy charge distributions exhibiting certain point group symmetries. We derive a general form of the electrostatic interaction energy of two permeable, arbitrarily charged spherical shells in the Debye-H\"{u}ckel approximation and apply it to the case of particles with icosahedral, octahedral, and tetrahedral inhomogeneous charge distributions. We analyze in detail how charge distribution symmetry modifies the interaction energy and find that local charge inhomogeneities reduce the repulsion of two overall equally charged particles, while sufficient orientational variation in the charge distribution can turn the minimum interaction energy into an attraction. Additionally we show that larger patches and thus lower symmetries and wave numbers result in bigger attraction given the same variation.
\end{abstract}


\maketitle

\section{Introduction}

Most commonly used theoretical models of virus capsids usually consider charge to be homogeneously distributed across one or two spherical shells representing the viral capsid shell~\cite{Siber2012,Lin2012,ALB2011,Prinsen2010,Zandi2006}. Reality is more complex, and the full 3D resolution of capsid structure shows significant variation in the distribution of charged amino acids across and along the capsid, carrying a signature of the underlying symmetries of the structure~\cite{Baker1999,ALB2012}. Local variations of charge on the capsid can play an important role in different scenarios, from RNA-capsid interactions~\cite{Belyi2006,Schoot2005}, interactions of the capsid with polyvalent ions~\cite{Zhao1997}, to capsid-capsid electrostatic interactions~\cite{Natarajan1998} or nanoparticle-templated assembly of virus-like particles~\cite{Daniel2010}.

In addition, electrostatic interactions can be the stabilizing force in ionic colloidal crystals of binary mixtures of oppositely charged particles which have been observed both in experiment and in simulation~\cite{Leunissen2005,Bartlett2005,Hynninen2006,Maskaly2006}. In such cases the assumed interaction potential is usually taken to be spherically symmetric, facilitating assembly but also severely limiting potential symmetries of assembled crystals~\cite{Juhl2006}. It is thus of special interest to investigate more general forms of the interaction potential between colloidal particles that are often anisotropic in shape and contain inhomogeneously distributed interacting moieties~\cite{Doye2007,Glotzer2007}. An important example are the patchy colloids as well as inverse patchy colloids with heterogeneously distributed interacting patches that can form a richer variety of admissible structures~\cite{Bianchi2011a,Bianchi2011,Noya2007,Doppelbauer2012}. Charged patchy colloids with heterogeneously charged surfaces belong to the same variety of complex colloidal particles except that in charged systems the interaction is usually much longer ranged than the standard bonding distances in ordinary patchy systems with limited valence~\cite{Bianchi2011a}.

Contrary to classical colloids, virus capsids, which are the main motivation for our work, offer a number of additional features that colloids do not possess. A few more prominent properties are the unique, spatially-defined chemistries on their surfaces as well as absolute monodispersity in size and mass distributions~\cite{Juhl2006}. Such monodisperse viral colloids are highly symmetrical~\cite{Baker1999,Salunke1989}, with pronounced symmetry-related variation in particle surface chemistry. These surface structural features make them a possible component material for the production of biophotonic crystals~\cite{Juhl2006,Parker2007}, non-close-packed crystalline structures (see Ref.~\onlinecite{Natarajan1998} and references therein) and virus-like nano particles that promise to invigorate the search for the perfect gene-therapy vector \cite{Steinmetz}.

{\em In vitro} the interviral interaction potential is characterized by weak long-range electrostatic repulsion and sterically mediated close-range attraction~\cite{Juhl2006}. Our focus in this work will reside on the electrostatic part, and we will derive an analytical (closed-form) expression for the interaction free energy between two spherical shells carrying arbitrary inhomogeneous surface charge distributions within the linearized Poisson-Boltzmann (Debye-H\"uckel) approximation~\cite{Andelman2006}. This result will enable us to analyze models of highly symmetric charge distributions mimicking the symmetries of viruses within the approach recently proposed by~\citet{Lorman2007,Lorman2008} in order to describe the mass distributions in spherical viruses.

The rest of the paper is structured as follows: In order to introduce the setting, we first summarize in Sec.~\ref{sec:oneshell} results obtained in previous works for the electrostatic self-energy of a single arbitrarily charged spherical shell in the Debye-H\"uckel (DH) approximation. Building on this we derive an analytical expression for the electrostatic interaction of two such shells in Sec.~\ref{sec:twoshell}. Appendices~\ref{app:app1} and~\ref{app:app2} contain the details of the derivation and certain limiting cases. We first use the derived expression to analyze a simple model of axially symmetric quadrupole surface charge distribution in Sec.~\ref{sec:quad}, comparing some of the results with those obtained in the literature for a similar case, but with a different model~\cite{Bianchi2011}. In Sec.~\ref{sec:sym} we then introduce more general surface charge densities with tetrahedral, octahedral, and icosahedral symmetry and use them to examine in more detail the electrostatic interactions between shells carrying such charge distributions. We end with a discussion and conclusions in Sec.~\ref{sec:con}.

\section{Derivation of interaction energy in Debye-H\"uckel approximation}

Throughout the paper we will deal with electrostatics of charged shells in a monovalent salt solution within the framework of the linearized mean-field Debye-H\"uckel (DH) theory~\cite{Andelman2006,Kjellander2008,Naji2005,Naji2010}. This approximation is reasonable for sufficiently small surface charge densities on the surface of the particles, low ion valencies, high medium dielectric constant, or high temperatures. All the requirements are in general well fulfilled for the case of monovalent salt solutions and the surface charge densities relevant for viruses or colloidal particles considered here.

\subsection{Self-energy of a charged spherical shell}\label{sec:oneshell}

We first briefly sum up the relevant parts in the derivation of the electrostatic self-energy of an arbitrarily charged spherical shell in a salt solution in order to introduce some concepts important for the rest of the paper. The radius of the shell is $R$ and the surface charge density on the shell can be expanded in terms of spherical harmonics,
\begin{equation}
\label{eq:scd}
\sigma(\Omega)=\sum_{l,m}\sigma(lm)Y_{lm}(\Omega).
\end{equation}
The requirement that the surface charge density be real imposes the constraint
$\label{eq:ilm}
\sigma^*(lm)=(-1)^m\sigma(l\underline{m}),
$
where we have introduced the notation $\underline{m}=-m$. The electrostatic potential $\varphi$ is given by the solution of the DH equation
\begin{equation}
\nabla^2\varphi=\kappa^2\varphi,
\label{DHequ}
\end{equation}
where $\kappa$ is the inverse DH screening length of a monovalent $1:1$ salt with bulk concentration $c_0$, $ \kappa=\sqrt{8\pi l_Bc_0}$, with $l_B=e_0^2/4\pi\varepsilon\varepsilon_0k_BT$ the Bjerrum length and $T$ the absolute temperature. The dielectric constant of water is taken to be $\varepsilon=80$.

The electrostatic potentials inside and outside the shell,
\begin{eqnarray}
\varphi_I(r,\Omega)&=&\sum_{l,m} a(lm)i_l(\kappa r)Y_{lm}(\Omega),~{\rm and}\\
\label{eq:singpot}
\varphi_{II}(r,\Omega)&=&\sum_{l,m} b(lm)k_l(\kappa r)Y_{lm}(\Omega),
\end{eqnarray}
are expanded in terms of spherical harmonics $Y_{lm}(\Omega)$ and the modified spherical Bessel functions of the first and second kind~\cite{Abramowitz1965},
\begin{eqnarray}
i_l(x)&=&\sqrt{\frac{\pi}{2x}}\,I_{l+1/2}(x)\quad\mathrm{and}\\
k_l(x)&=&\sqrt{\frac{\pi}{2x}}\,K_{l+1/2}(x).
\end{eqnarray}
The potential has to be continuous on the surface of the shell, $\varphi_I(R,\Omega)=\varphi_{II}(R,\Omega)$ for all values of $\Omega$, and the surface charge on the shell gives rise to a discontinuity of the electrostatic potential on its surface in the standard form
\begin{equation}
\left.\frac{\partial\varphi}{\partial r}\right|_{r=R_-}-\left.\frac{\partial\varphi}{\partial r}\right|_{r=R_+}=\frac{\sigma(\Omega)}{\varepsilon\varepsilon_0}.
\end{equation}
By applying the boundary conditions we find that the solutions for the potential have the same symmetry as the underlying surface charge density,
\begin{eqnarray}
\varphi_I(r,\Omega)&=&\sum_{l,m}\frac{C_0(l,\kappa R)}{\kappa\varepsilon\varepsilon_0}\frac{i_l(\kappa r)}{i_l(\kappa R)}\sigma(lm)Y_{lm}(\Omega),\\
\varphi_{II}(r,\Omega)&=&\sum_{l,m}\frac{C_0(l,\kappa R)}{\kappa\varepsilon\varepsilon_0}\frac{k_l(\kappa r)}{k_l(\kappa R)}\sigma(lm)Y_{lm}(\Omega).
\label{gfyew}
\end{eqnarray}
Here, we have defined
\begin{equation}
\label{eq:cnula}C_0(l,x)=x\,I_{l+1/2}(x)K_{l+1/2}(x).
\end{equation}
The electrostatic free energy in the DH limit can then be calculated as~\cite{Marzec1993,Verwey}
\begin{equation}
F_{DH}=\frac{1}{2}\oint_{\partial V}\sigma(\Omega)\varphi(R,\Omega)\mathrm{d}S.
\end{equation}
Using the orthogonality relations for the spherical harmonics together with the requirement that the charge density be real, we then obtain 
\begin{eqnarray}
\label{eq:singfree}
\nonumber F_{DH}&=&\frac{R^2}{2\kappa\varepsilon\varepsilon_0}\frac{4\pi\sigma_0^2}{1+\coth\kappa R}+\\
&+&\sum_{l>0}\frac{R^2}{2\kappa\varepsilon\varepsilon_0}C_0(l,\kappa R)\sum_m|\sigma(lm)|^2.
\end{eqnarray}
This result was already obtained by~\citet{Marzec1993}; the first term ($l=0$) is simply the self-energy of a uniformly charged shell in a salt solution~\cite{Siber2007}. Details of the derivation can be found in e.g. Ref.~\onlinecite{ALB2011}, where it was used to derive the free energy of a partially formed shell -- a special case where the surface charge density is a Heaviside step function in the azimuthal angle.

\subsubsection{Rotations of the shell(s)}

Given the original surface charge distribution of a shell [Eq.~\eqref{eq:scd}] in its reference frame, the expansion coefficients will in general change with different orientations of the shell as we rotate it. The easiest way to incorporate the rotations is through the Wigner matrices $D_{m'm}^{(l)}(\boldsymbol{\omega})$ given in terms of the Euler angles $\boldsymbol{\omega}=(\alpha,\beta,\gamma)$ in the $zyz$ notation:
\begin{equation}
D_{m'm}^{(l)}(\boldsymbol{\omega})=e^{-\imath m'\gamma}d_{m'm}^{\,l}(\beta)e^{-\imath m\alpha},
\end{equation}
where we use the definition of the small d-matrices $d_{m'm}^{\,l}(\beta)$ as given by~\citet{Rose1957}. From here the expansion coefficients of the rotated distribution can be expressed in terms of the original ones as~\cite{Rose1957,Gray}
\begin{equation}
\sigma'(lm)=\sum_{m'}D_{m'm}^{(l)}(\boldsymbol{\omega})\sigma(lm');
\end{equation}
the rotations do not mix different wave numbers $l$.

Due to the way the expansion coefficients feature in the self-energy of the shell [Eq.~\eqref{eq:singfree}] changing the orientation of the shell does not influence the result. However, when we will next consider the interaction of two such shells, different orientations of both shells will play a role. In the rest of the paper we will thus denote the coefficients of the original surface charge distributions of the two shells in their respective reference frames by $\sigma(lm,i)$ and those of the rotated distributions by $\sigma'(lm,i)$, while the orientations of the two shells will be given by the Euler angles $\boldsymbol{\omega}_i$.

\subsection{Interaction free energy of two arbitrarily charged spherical shells}\label{sec:twoshell}

Taking two charged shells as studied in the previous Section, we now derive the expression for their interaction energy. The basic aspects of the method were already presented by ~\citet{Langbein} in the calculation of van der Waals forces between molecules, and similar approaches have been used to calculate the force and interaction energies of charged shells/colloids in simpler~\cite{Glendinning1983,Carnie1993,Ohshima1993,Ohshima1995,Ohshima1996} and more involved~\cite{Hoffmann2004,Linse2008,Bichoutskaia2010,Bianchi2011} scenarios.

\begin{figure}[!htb]
\includegraphics[width=8.5cm]{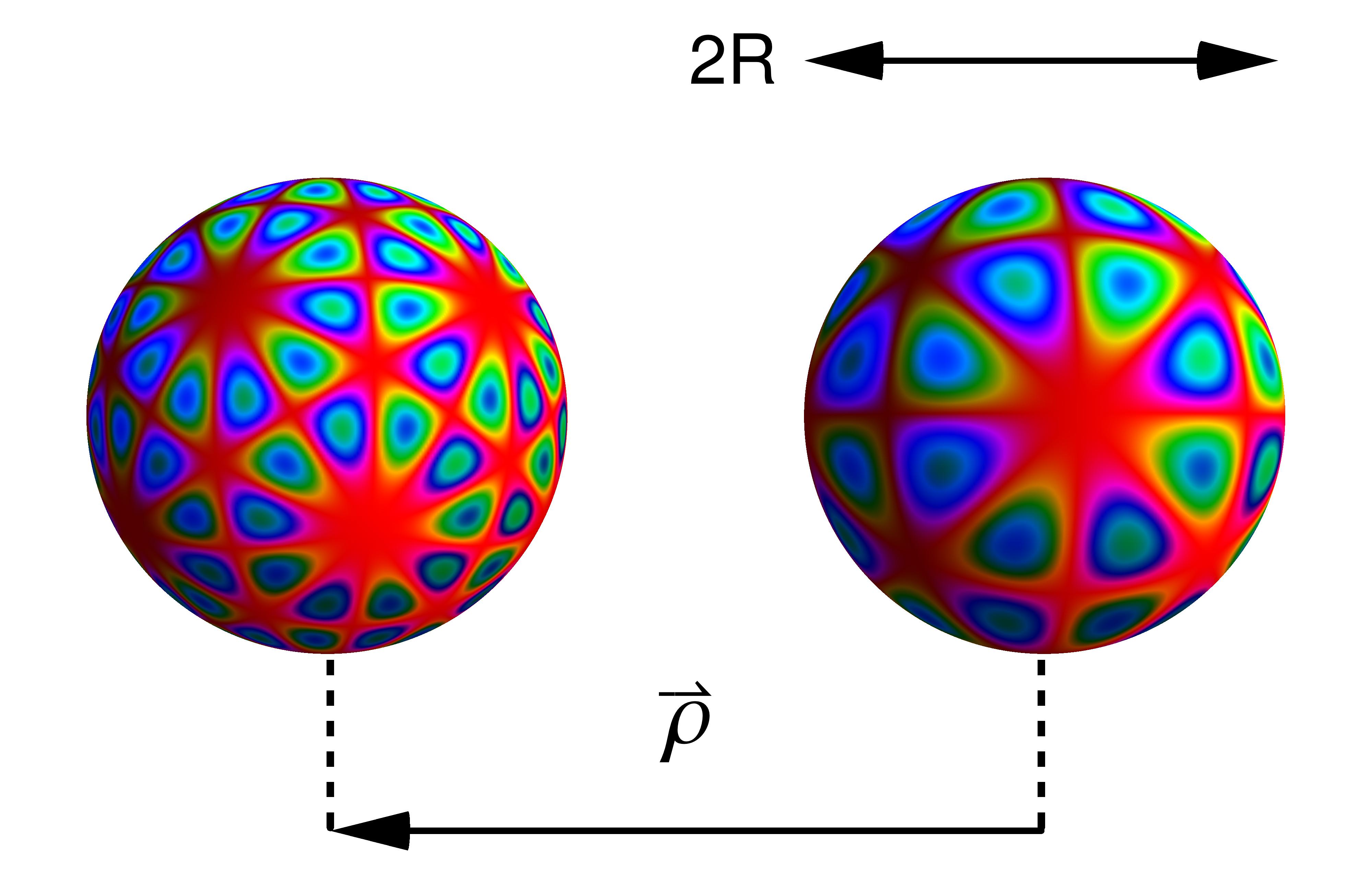}
\caption{\label{fig:sketch}The system under consideration: two shells with arbitrary surface charge densities $\sigma^{(j)}(\Omega)$ (in this case, with icosahedral and octahedral symmetry) and radii $R$ are located at a distance $\rho$. Either or both of the distributions can be additionally rotated, with the orientations given by the Euler angles $\boldsymbol{\omega}_j$. The shells are in a monovalent salt solution with bulk concentration $c_0$ giving rise to the electrostatic screening length $\kappa^{-1}$. The background color of the shells represents the homogeneous charge contribution, whereas the other tones depict the local patches of charge of opposing signs.}
\end{figure}

The two shells under consideration carry surface charge densities $\sigma^{(j)}(\Omega)$ expanded as in Eq.~\eqref{eq:scd}; the orientations of the two shells can be arbitrary and are given by the Euler angles $\boldsymbol{\omega}_j=(\alpha_j,\beta_j,\gamma_j)$. The shells and their respective properties are denoted by indices 1 and 2. The radii of the shells are $R_j=R$, and the distance between their centers is $r_{21}=\rho$. The vector $\boldsymbol{\rho}=\mathbf{r}_2-\mathbf{r}_1=(\rho,\xi,\eta)$ in spherical coordinates connects the two coordinate systems. The rest of the system properties are the same as in Sec.~\ref{sec:oneshell}; Fig.~\ref{fig:sketch} shows a sketch of the system.

\subsubsection{Electrostatic potential \ldots}

The solution of the DH equation for the mean electrostatic potential [Eq.~\eqref{DHequ}] can be written as a linear superposition $\psi=\psi_1+\psi_2$, where $\psi_j$ is the electrostatic potential of a single ($j$-th) shell. We can divide the system into three regions: inside the first shell ($\psi_{in,1}$), inside the second shell ($\psi_{in,2}$), and outside both shells ($\Psi$). The boundary conditions are again the continuity of the potential at the surface of each shell,
\begin{equation}
\label{eq:bc1}\psi_{in,j}(R_j^-,\Omega)=\Psi(R_j^+,\Omega),
\end{equation}
and the discontinuity of the derivative, which has to be proportional to the charge on the shell,
\begin{equation}
\label{eq:bc2}\left.\frac{\partial\psi_{in,j}}{\partial r_j}\right|_{r_j=R_j^-}-\left.\frac{\partial\Psi}{\partial r_j}\right|_{r_j=R_j^+}=\frac{\sigma^{(j)}(\Omega)}{\varepsilon\varepsilon_0}.
\end{equation}

The electrostatic potential can be expanded in terms of modified spherical Bessel functions of the first and second kind and spherical harmonics. We will follow the notation introduced in Ref.~\onlinecite{Clercx1993} and define
\begin{eqnarray}
\psi_{lm}^+(\mathbf{r})&=&i_l(\kappa r)Y_{lm}(\Omega)\quad\mathrm{and}\\
\psi_{lm}^-(\mathbf{r})&=&k_l(\kappa r)Y_{lm}(\Omega).
\end{eqnarray}
Thus we can write for the solution inside each shell
\begin{equation}
\psi_{in,j}(\mathbf{r}_j)=\sum_{l,m}a(lm,j)\psi_{lm}^+(\mathbf{r}_j),
\end{equation}
and the solution at a point $\mathbf{r}$ outside both shells
\begin{equation}
\Psi(\mathbf{r})=\sum_{l,m}\left[b(lm,1)\psi_{lm}^-(\mathbf{r}-\mathbf{r}_1)+b(lm,2)\psi_{lm}^-(\mathbf{r}-\mathbf{r}_2)\right]
\end{equation}
due to the linearity of the DH equation. We will also use the notation $\psi_{lm}^\pm(1)$ and $\psi_{lm}^\pm(2)$ for the potentials written in the coordinate systems of the first and the second shell, respectively.

To move between the two coordinate systems we have to use an addition theorem for the solutions of the modified Helmholtz equation in spherical coordinates~\cite{Langbein}. \citet{Clercx1993} give the theorem for an arbitrary relative positioning of the two shells specified by the vector $\boldsymbol{\rho}$. However, since all possible orientations of the two shells are already taken into account via rotations given by the Euler angles $\boldsymbol{\omega}_j$ we can fix their relative position to $\boldsymbol{\rho}=\rho\,\hat{\mathbf{z}}$. This allows us to write a simplified version of the addition theorem from Ref.~\onlinecite{Clercx1993},
\begin{equation}
\psi_{lm}^-(1)=\sum_{p,q}{\cal L}_{lm}^{pq}(1)\psi_{pq}^+(2),
\end{equation}
where the function ${\cal L}_{lm}^{pq}(1)$ is expressed in terms of modified Bessel functions of the second kind and Wigner 3-j symbols:
\begin{eqnarray}
\label{eq:llm}
\nonumber
{\cal L}_{lm}^{pq}(1)&\equiv&\sum_{s}(-1)^{p+m}(2s+1)\sqrt{(2l+1)(2p+1)}\times\\
&\times&k_s(\kappa\rho)\left(\begin{array}{ccc}
      l & s & p \\
      0 & 0 & 0
      \end{array}
\right)
\left(\begin{array}{ccc}
      l & s & p \\
      \underline{m} & 0 & q
      \end{array}
\right).
\end{eqnarray}
The theorem in the opposite direction, $\psi_{lm}^-(2)=\sum_{p,q}{\cal L}_{lm}^{pq}(2)\psi_{pq}^+(1)$, is essentially the same, with an additional factor of $(-1)^s$ appearing in the summation over $s$ in the function ${\cal L}_{lm}^{pq}(2)$. Some of the useful properties of the function ${\cal L}_{lm}^{pq}$ are listed in Appendix~\ref{app1:sec1}.

Combining now the two boundary conditions in Eqs.~\eqref{eq:bc1} and~\eqref{eq:bc2} on the surface of e.g. the second shell, we obtain the coefficients of the solution for the potential in the exterior and interior,
\begin{eqnarray}
\label{eq:blm2} b(lm,2)&=&\frac{\sigma(lm,2)}{\kappa\varepsilon\varepsilon_0}\frac{C_0(l,\kappa R)}{k_l(\kappa R)},\\
\nonumber
\label{eq:alm2} a(lm,2)&=&\frac{\sigma(lm,2)}{\kappa\varepsilon\varepsilon_0}\frac{C_0(l,\kappa R)}{i_l(\kappa R)}+\\
&+&\sum_{p,q}\frac{\sigma(pq,1)}{\kappa\varepsilon\varepsilon_0}\frac{C_0(p,\kappa R)}{k_p(\kappa R)}{\cal L}_{pq}^{lm}(1),
\end{eqnarray}
where the function $C_0(l,x)$ is defined in Eq.~\eqref{eq:cnula}. The expressions for the coefficients of the first shell are similar with $1\leftrightarrow 2$ exchanged everywhere. The procedure for obtaining the coefficients is given in Appendix~\ref{app1:sec2}. This completes the solution for the electrostatic potential inside and outside the shells.

\subsubsection{\ldots and the interaction energy}

From the solution for the DH potential we can calculate the free energy of the two shells as $F_{el}=F_{el,1}+F_{el,2}$ where, similarly as in the case of a single shell,
\begin{equation}
\label{eq:felj}F_{el,j}=\frac{1}{2}\oint_{\partial V_j}\sigma^{(j)}(\Omega_j)\,\psi(r_j=R,\Omega_j)\,\mathrm{d}S_j,
\end{equation}
where now $\psi(r_j=R,\Omega_j)$ designates the potential of both shells so that their electrostatic interaction energy at a distance $\rho$ is then given as~\cite{Ohshima1995,Kornyshev1997}
\begin{equation}
V_{int}(\rho)=F_{el}(\rho)-F_{el}(\infty).
\end{equation}

First we note that the expansion coefficients $a(lm,2)$ [Eq.~\eqref{eq:alm2}] can be written in the form
\begin{equation}
a(lm,2)=a_0(lm,2)+\sum_{p,q}b(pq,1){\cal L}_{pq}^{lm}(1),
\end{equation}
with $a_0(lm,2)$ being the expansion coefficients one obtains in the case of a single shell (Sec.~\ref{sec:oneshell}). By writing out in full both the surface charge distribution and the potential at the shell surface in Eq.~\eqref{eq:felj} we can then use the properties of spherical harmonics to write the free energy of the second shell in the form
\begin{equation}
F_{el,2}=F_{el,2}^{(0)}+\frac{R^2}{2}\sum_{l,m}\sum_{p,q}b(pq,1)\sigma^*(lm,2){\cal L}_{pq}^{lm}(1)i_l(\kappa R),
\end{equation}
where $F_{el,2}^{(0)}$ is now the free energy in the case of a single, isolated shell [Eq.~\eqref{eq:singfree}].

In the interaction energy, these terms cancel out with the contribution of the two shells at infinite separation, and we are left with
\begin{eqnarray}
\nonumber V_{int}(\rho)&=&\frac{R^2}{2}\sum_{l,m}\sum_{p,q}\Big[b(pq,1)\sigma^*(lm,2){\cal L}_{pq}^{lm}(1)+\\
&+&b(pq,2)\sigma^*(lm,1){\cal L}_{pq}^{lm}(2)\Big]i_l(\kappa R).
\end{eqnarray}
We can also write this expression in terms separated by their respective wave numbers $l$ and $p$, and define
\begin{equation}
\label{eq:vint}V_{int}=\sum_lV_{ll}+\sum_{l\neq p}(V_{lp}+V_{pl})\equiv\frac{1}{2}\sum_lW_{ll}+\sum_{l> p}W_{lp}.
\end{equation}
The functions $W_{lp}$ can be further simplified using some of the properties of Wigner 3-j symbols and Bessel functions (see the Appendix~\ref{app1:sec3}), finally leading to the interaction energy terms
\begin{widetext}
\begin{eqnarray}
\label{eq:wlp}
\nonumber W_{lp}&=&\frac{R^2}{\kappa\varepsilon\varepsilon_0}C(l,p,\kappa R)\sum_{m,s}(-1)^{l+m}\,\Re\Big[\sigma(lm,1)\sigma^*(pm,2)+(-1)^{l+p}\sigma(pm,1)\sigma^*(lm,2)\Big]\times\\ &\times&(2s+1)\sqrt{(2l+1)(2p+1)}\,k_s(\kappa\rho)\left(\begin{array}{ccc}
      l & s & p \\
      0 & 0 & 0
      \end{array}
\right)
\left(\begin{array}{ccc}
      l & s & p \\
      \underline{m} & 0 & m
      \end{array}
\right),
\end{eqnarray}
\end{widetext}
where we have defined
\begin{equation}
\label{eq:aux}C(l,p,x)=x\,I_{l+1/2}(x)I_{p+1/2}(x).
\end{equation}
The entire dependence on the orientations of the two shells relative to their respective reference frames is hidden in the expansion coefficients $\sigma(lm,i)$, and upon replacing them with the rotated ones, $\sigma'(lm,i)$, we get the interaction free energy terms $W_{lp}(\rho,\boldsymbol{\omega}_1,\boldsymbol{\omega}_2)$.

\section{Simple case: axially symmetric quadrupole distribution}\label{sec:quad}

With the expression for the interaction free energy known, we can now analyze different scenarios. Firstly, we will shortly discuss an example of two shells carrying axially symmetric quadrupole distributions, a case similar to the one considered by~\citet{Bianchi2011} for inverse patchy colloids. There, however, the colloids are impermeable to salt ions, unlike the shells considered here.

The model can be described by a surface charge distribution of the form
\begin{equation}
\sigma(\Omega)=\sigma_2Y_{20}(\Omega)+\sqrt{4\pi}\sigma_0Y_{00}(\Omega),
\end{equation}
which is axially symmetric around the $z$ axis and has two patches of the same charge at the poles and a patch of opposite charge around the equator. By varying $\sigma_0$ we can also include a homogeneous background, where we set the coefficient $\sigma(00,i)=\sqrt{4\pi}\sigma_0$ to ensure that the total charge is given by $Q=4\pi R^2\sigma_0$. Due to the symmetry the rotation angles are limited and we need consider only the azimuthal angles of the two shells, $\beta_i$.

\begin{figure}[!htb]
\includegraphics[width=8.5cm]{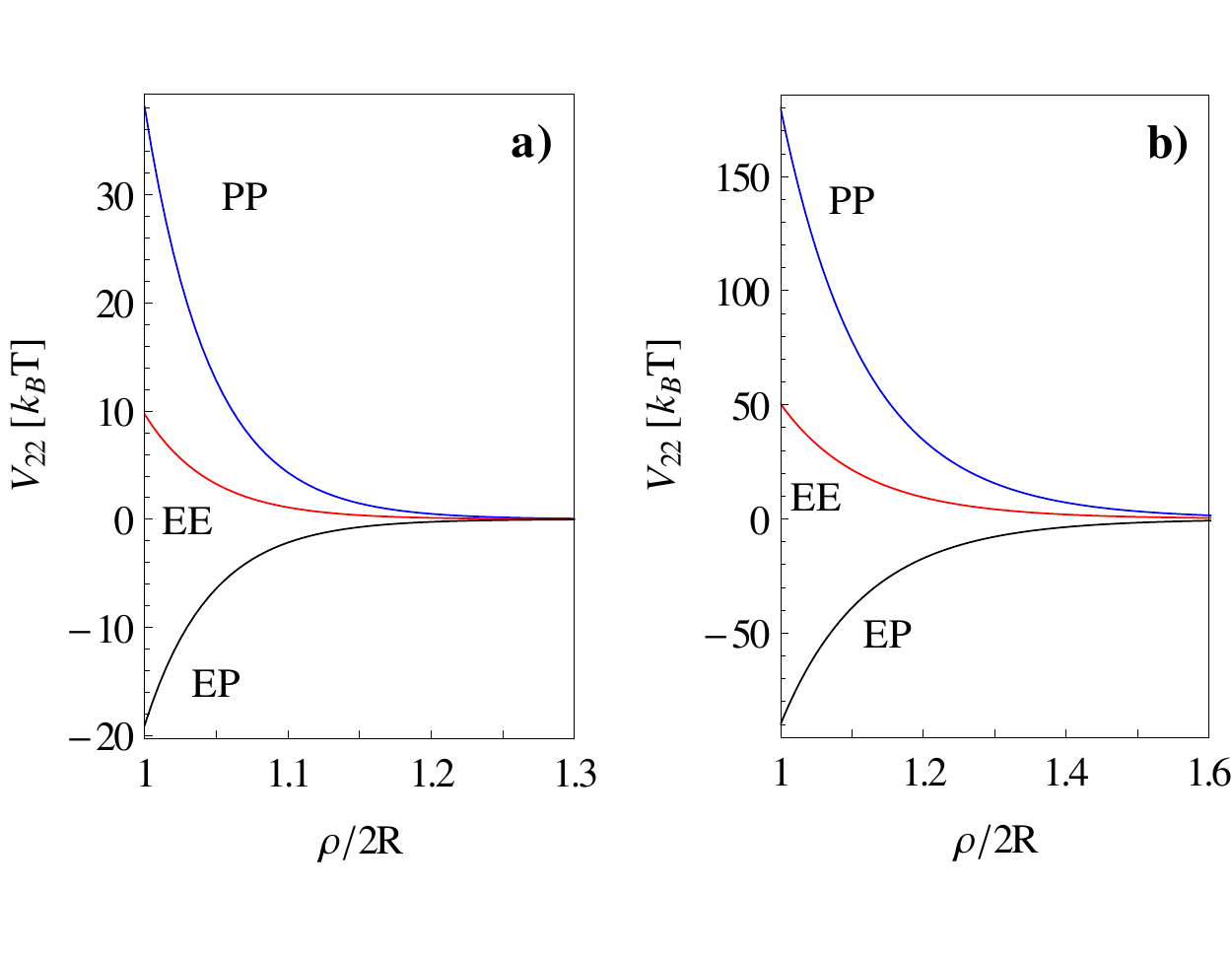}
\caption{Quadrupole-quadrupole interaction energy $V_{22}$ as a function of the intershell separation for three different configurations of the shells, pole-pole (PP), equator-pole (EP), and equator-equator (EE). The radii of the shells are $R=10$ nm, with {\bf a)} $\kappa R\approx10$ and {\bf b)} $\kappa R\approx3$, and the strength of the variation is $\sigma_2=1$ $e_0/\mathrm{nm}^2$.}
\label{fig:v22_radial}
\end{figure}

Focusing initially only on the quadrupole-quadrupole interaction energy $V_{22}(\rho,\boldsymbol{\omega}_i)$ we can see from the symmetry of the problem that the most optimal arrangement of the shells in this case is equator-pole (EP). Thus, we consider three orientational configurations (EE, EP, PP) and plot the radial dependence of the interaction energy for the three cases in Fig.~\ref{fig:v22_radial}. Similar to observations in Ref.~\onlinecite{Bianchi2011} we find that the interaction energy is most repulsive in the PP configuration, is smaller for the EE configuration, and turns into attraction in the EP case. With increasing $\kappa R$ the scale of the interaction energy diminishes, and the interaction falls of more quickly with increasing inter-shell separation. The interaction between the shells also remains either purely repulsive or purely attractive with respect to the separation between them.

\begin{figure}[!htb]
\includegraphics[width=7.5cm]{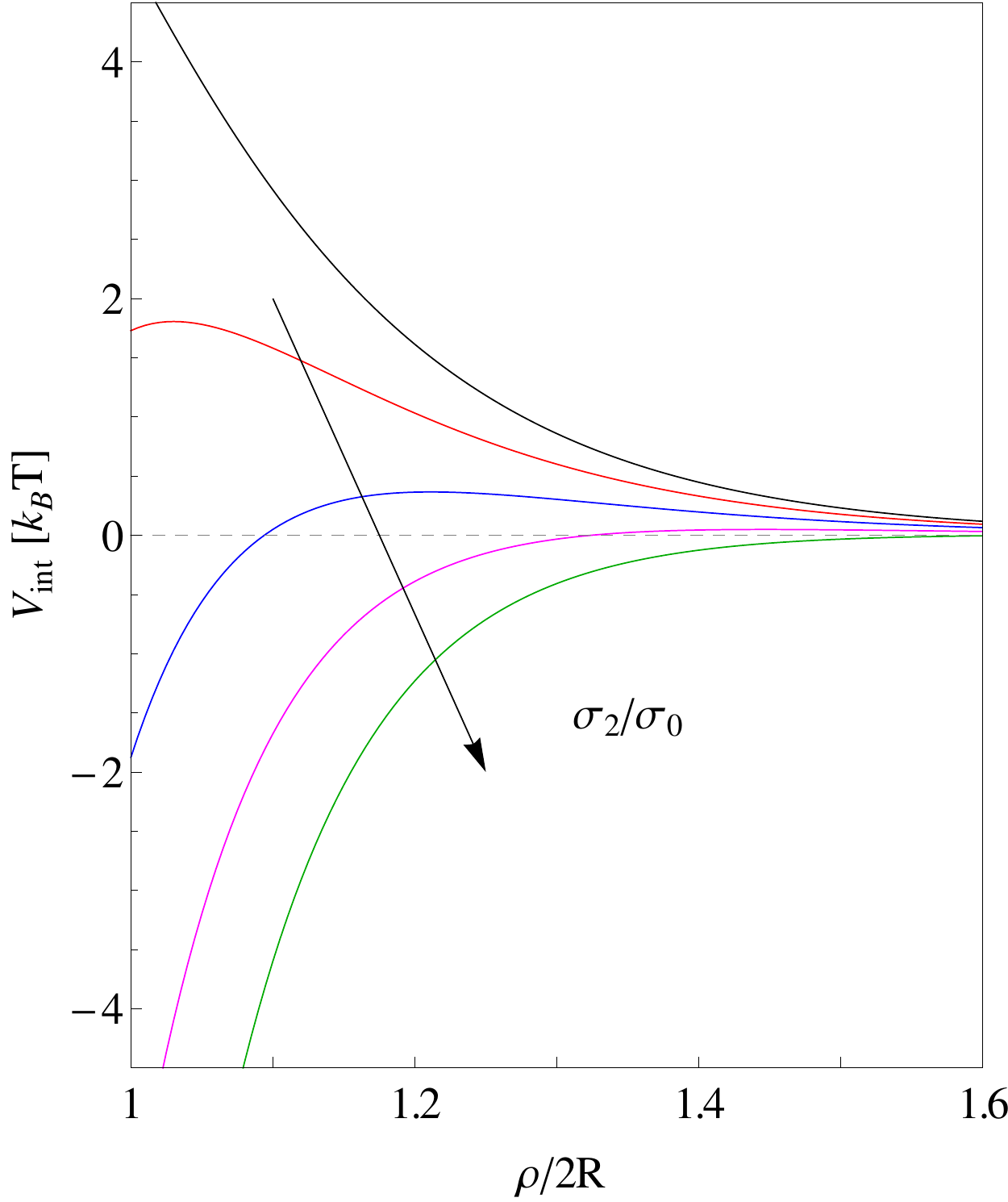}
\caption{Total interaction energy of two charged shells with quadrupole variation [Eq.~\eqref{eq:vint_quad}] as a function of the intershell distance for different ratios of the quadrupole and homogeneous contributions to the surface charge density, $\sigma_2/\sigma_0\in[0.45,0.65]$. Radii of the shells are $R=10$ nm with $\kappa R\approx3$, and the shells are in the EP configuration.}
\label{fig:quad_ratio}
\end{figure}

By adding a homogeneous background to the two shells we obtain two additional contributions to the interaction energy,
\begin{equation}
\label{eq:vint_quad}
V_{int}(\rho,\boldsymbol{\omega}_i)=V_0(\rho)+W_{20}(\rho,\boldsymbol{\omega}_i)+V_{22}(\rho,\boldsymbol{\omega}_i).
\end{equation}
The relative contributions of the terms are governed by the ratio $\sigma_2/\sigma_0$, and Fig.~\ref{fig:quad_ratio} shows how the total interaction energy changes with respect to this ratio for the EP configuration of the shells. The homogeneous contribution to surface charge density $\sigma_0=0.1$ $e_0/\mathrm{nm}^2$ is kept fixed and amounts to a charge of $Q\approx 125$ $e_0$ on a shell of radius $R=10$ nm, while we vary the quadrupole component of charge on the shells. The homogeneous background (of the same sign on both shells) adds an additional repulsive contribution  which prevails for small variations in surface charge density; thus, even the EP configuration of the shells results in a repulsive interaction. However, upon increasing the quadrupole part the interaction becomes attractive for larger and larger distances, with the repulsive contribution slowly vanishing.

\begin{figure*}[!htb]
\includegraphics[width=17cm]{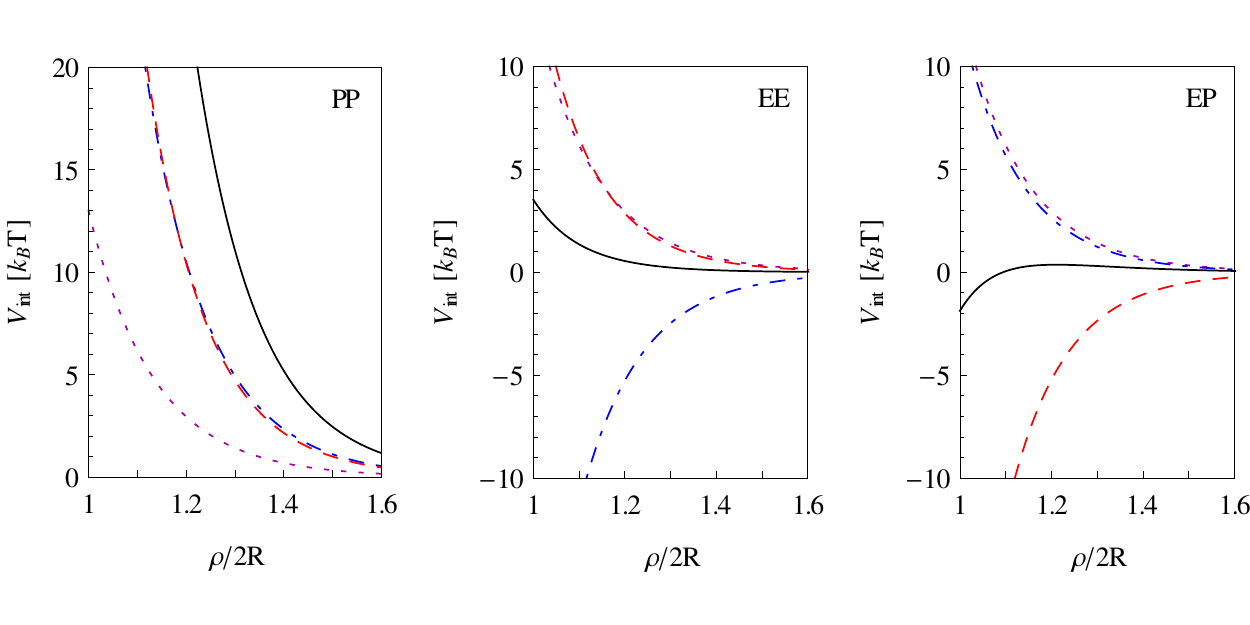}
\caption{Total interaction energy of two charged shells with quadrupole variation $V_{int}$ (full line) and the relative contributions ($V_0$ -- dotted line, $W_{20}$ -- dot-dashed line, $V_{22}$ -- dashed line) as a function of the intershell separation for three different orientations. As before, the radii of the shells are $R=10$ nm with $\kappa R\approx3$, with the ratio $\sigma_2/\sigma_0=0.55$.}
\label{fig:quad_radial}
\end{figure*}

Figure~\ref{fig:quad_radial} shows the contribution of the different terms in the interaction free energy [Eq.~\eqref{eq:vint_quad}] for the PP, EE, and EP orientations of the two shells. The homogeneous term $V_0$ contributes only a repulsive component regardless of their orientations, whereas the cross-term $W_{20}$ changes the behavior of the interaction energy with respect to the shell orientations. In the PP configuration every component is repulsive, totalling to an even bigger repulsion than in the simpler quadrupole-quadrupole case. In the EP configuration, the total interaction is attractive due to the quadrupole-quadrupole term, the cross-term being consistently repulsive. The cross-term becomes attractive in the EE configuration, where the quadrupole-quadrupole term is now repulsive; the latter, together with the homogeneous contribution, prevails and the total interaction is repulsive.

\section{Interaction between two shells with symmetric charge distributions}\label{sec:sym}

In this Section we will examine electrostatic interactions between model virus-like particles carrying surface charge distributions with icosahedral symmetry; we will also include octahedral and tetrahedral symmetries for comparison. In order to do this, we next show how to obtain distribution functions invariant under a given symmetry group.

\subsection{Lower symmetry charge distributions: tetrahedral, octahedral, and icosahedral symmetry}\label{sym:chr}

In constructing the surface charge distribution function possessing a given symmetry, we will use the ideas presented in Refs.~\onlinecite{Lorman2007} and~\onlinecite{Lorman2008}, with appropriately described details. The sought-for symmetry adapted functions are obtained by considering crystallization on a sphere with symmetry reduction from the (isotropic) 3D rotation group $SO(3)$ to the icosahedral (octahedral, tetrahedral) point subgroup ${\cal Y}\subset SO(3)$~\cite{Elliott1986}.

The symmetry adapted icosahedral functions are thus obtained by reducing the irreducible representations $D^{(l)}$ of $SO(3)$ to the icosahedral point group ${\cal Y}$. We stipulate that the identical representation of the point group appears at least once in the reduction so that the resulting structure is invariant under all symmetry operations of the icosahedral group. This restricts the allowed wave numbers $l$ to~\cite{Lorman2007,Lorman2008}
\begin{equation}
l_{ico}=6i+10j\,(+15).
\end{equation}
Odd $l$s lack inversion symmetry, making such distributions consistent with the asymmetry of the proteins which constitute the viral capsid~\cite{Lorman2008}. Even $l$s are included for comparison and the corresponding distributions fit a dodecahedral geometry better than the icosahedral. For octahedral and tetrahedral groups the same procedure yields the allowed wave numbers $l_{oct}=4i+6j\,(+9)$ and $l_{tet}=4i+6j\,(+3)$.

\begin{figure*}
\includegraphics[width=17cm]{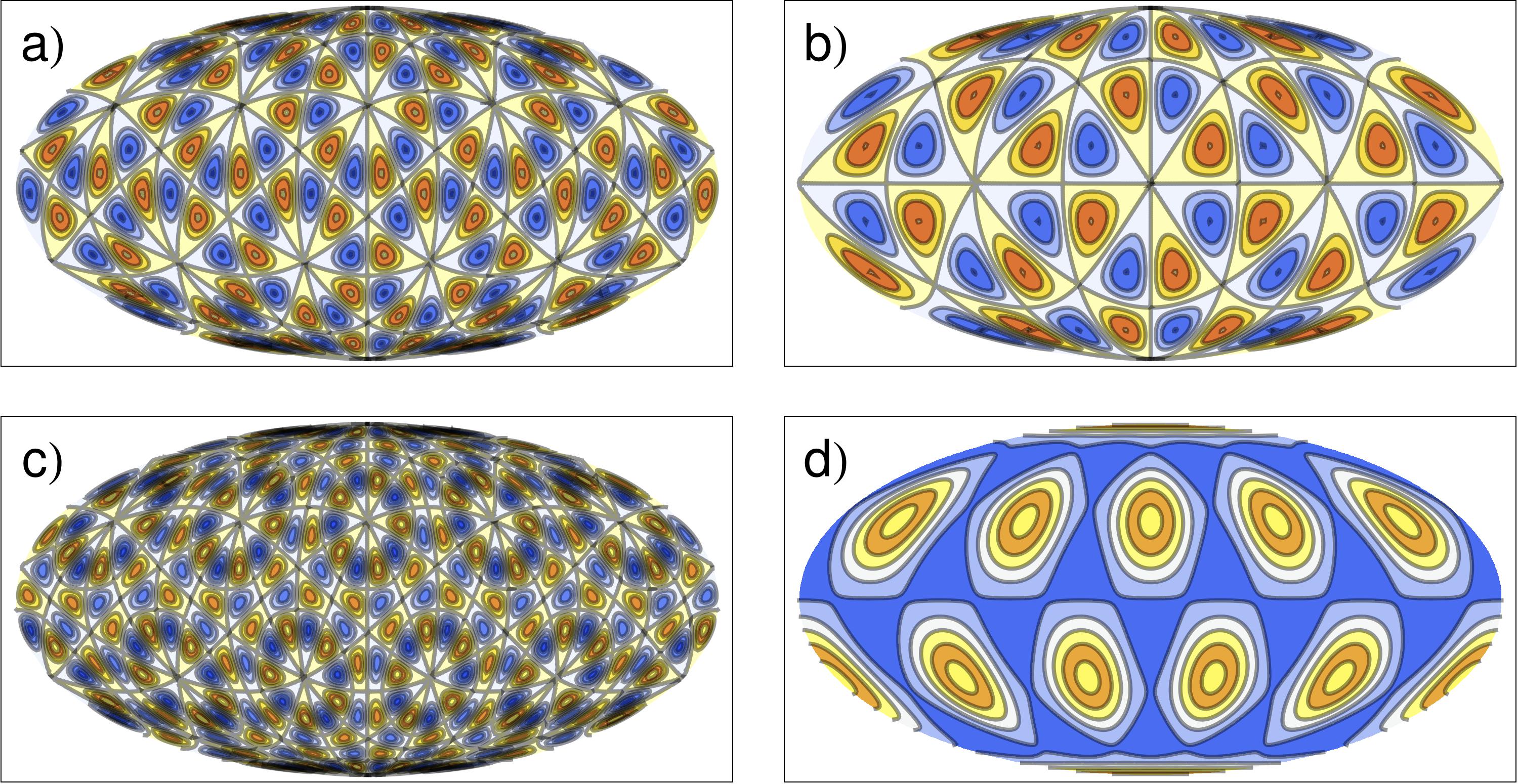}
\caption{\label{fig:distributions}Symmetry adapted irreducible density functions $f_l(\Omega)$ for different symmetries and wave numbers: {\bf a)} Icosahedral symmetry with $l=15$, {\bf b)} octahedral symmetry with $l=9$, {\bf c)} icosahedral symmetry with $l=21$, and {\bf d)} dodecahedral symmetry with $l=6$. The distributions are mapped from a sphere to a plane using the Mollweide projection~\cite{Snyder1987}. Warmer colors correspond to regions with positive values and colder colors correspond to regions with negative values.}
\end{figure*}

For a selected wave number the explicit form of the surface charge density $\sigma_l(\Omega)$ is then given by the basis functions $f_l^k(\Omega)$, $k=1,\ldots,m_A(l)$, of all $m_A$ totally symmetric representations of the icosahedral group $\cal{Y}$ in the restriction of the ``active'' irreducible representations of the $SO(3)$. For lower wave numbers there is additionally only one identical representation for each allowed wave number, $m_A(l)=1$, and we have
\begin{equation}
\sigma_l(\Omega)=\sigma_lf_l(\Omega).
\end{equation}
The explicit form of the symmetry adapted irreducible icosahedral density function $f_l(\Omega)$ for a given wave number is obtained by averaging the spherical harmonics over the icosahedral symmetry group. Thus we obtain distributions of charge on a sphere which have icosahedral symmetry, characterized by an allowed wave number $l$:
\begin{equation}
\sigma_l(\Omega)=\sigma_l\sum_m I_{lm}Y_{lm}(\Omega),
\end{equation}
where the expansion coefficients are obtained with the averaging procedure as
\begin{equation}
I_{lm}=\frac{1}{60}\sum_{G\in\cal{Y}}D_{mm'}^{(l)}\Big(G(\alpha,\beta,\gamma)\Big).
\end{equation}
Here, $G(\alpha,\beta,\gamma)$ are the symmetry operations of the icosahedral group given in terms of the Euler angles and $D_{mm'}^{(l)}$ are the Wigner matrices~\cite{Rose1957}. The irreducible density functions for the octahedral and tetrahedral group are obtained in an analogous fashion. A few examples of the irreducible density functions $f_l$ of the symmetry groups are shown in Fig~\ref{fig:distributions}.

It is important to stress that the expansion coefficients $I_{lm}$ of the irreducible density functions $f_l(\Omega)$ are completely determined by the required invariance under the point group and the only free parameter left is $\sigma_l$. The coefficients are purely real for even $l$ and purely imaginary for odd $l$, with an additional useful property that $\sum_m|I_{lm}|^2=1$, i.e. the functions $f_l$ are normalised. As can be seen in Fig.~\ref{fig:distributions}, the number of local minima and maxima increases with increasing wave number; for the icosahedral symmetry, their number can be correlated with the Caspar-Klug triangulation number (cf. Ref.~\onlinecite{Lorman2008}). Additionally, the functions with odd $l$ have equal extrema, while those with even $l$ differ in the depth and shape of the minima and maxima.

In what follows we will deal with surface charge distributions of the form
\begin{equation}
\label{eq:icochr}
\sigma(\Omega)=\sigma_0+\sigma_l\sum_mI_{lm}Y_l^m(\Omega),
\end{equation}
where the wave number $l$ and the symmetry fully determine the expansion coefficients. The total charge on the shell is $Q=4\pi R^2\sigma_0$ and the $\sigma_l$ gives only the variation of the charge due to the symmetry, contributing nothing to the total charge ($\delta_{\sigma}^2=\langle\sigma^2\rangle-\langle\sigma\rangle^2=\sigma_l^2$). We can connect the variation strength $\sigma_l$ with the symmetry, if we demand that the charge on the shell is of the same sign everywhere,
\begin{equation}
\label{eq:scrit}
\widetilde{\sigma}_lf_l(\Omega)+\sigma_0\geqslant 0\quad\forall\Omega,
\end{equation}
from where we obtain the critical variation strength $\widetilde{\sigma}_l=\sigma_0/|f_l^{min}|$.

\subsection{Self-energy}

From Eq.~\eqref{eq:singfree} and the aforementioned property of the functions $f_l(\Omega)$ that $\sum_m|I_{lm}|^2=1$, we have for the self-energy of a single shell with surface charge distribution with wave number $l$
\begin{equation}
F_{DH}^{(l)}=\frac{R^2\sigma_l^2}{2\kappa\varepsilon\varepsilon_0}C_0(l,\kappa R).
\end{equation}
Even though the symmetry has an imprint on the functions $f_l$ and consequently on the electrostatic potential, the free energy depends only on the wave number and the strength of the variation. The function $C_0(l,x)$ tends to $1/2$ in the limit when $x\to\infty$, regardless of the wave number. However, in the opposite limit of $x\to0$ it tends to zero as $x/(2l+1)$ in the lowest order of approximation.

\begin{figure}[!htb]
\includegraphics[width=8.5cm]{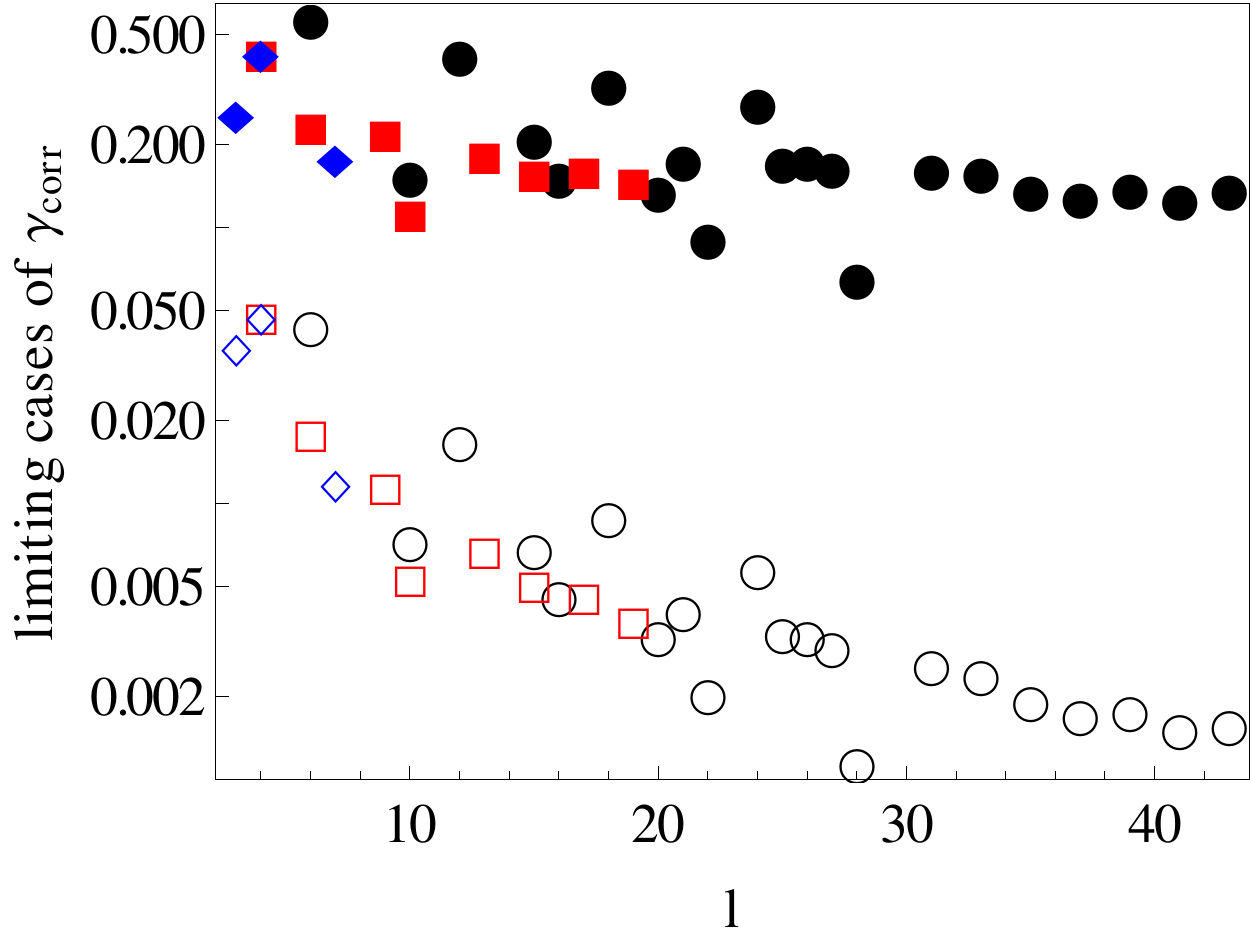}
\caption{Correction to the free energy of a charged shell [Eq.~\eqref{eq:fcorr2}] due to local charge variation characterized by wave number $l$. The strength of the variation is such that the charge on the shell is everywhere of equal sign, $\widetilde{\sigma}_l=\sigma_0/|f_l^{min}|$. Full symbols show the limit where $\kappa R\to\infty$ and empty symbols show the limit of $\kappa R\to 0$. Circles denote icosahedral, squares octahedral, and diamonds tetrahedral symmetry.}
\label{fig:limits}
\end{figure}

Upon the addition of homogeneously distributed background charge, the self-energy of such a shell can be written as
\begin{equation}
\label{eq:fcorr}F_{DH}=F_{DH}^{(0)}\left[1+\frac{1}{4\pi}\left(\frac{\sigma_l}{\sigma_0}\right)^2\frac{C_0(l,\kappa R)}{C_0(0,\kappa R)}\right].
\end{equation}
The ratio $C_0(l,x)/C_0(0,x)$ goes to 1 in the limit of $x\to\infty$, and falls off to an $l$-dependent constant in the limit of $x\to0$ as $(1+x)/(2l+1)$. Consequently, in the limit of $\kappa R\gg 1$ only the strength of the variation $\sigma_l$ and the ratio $\sigma_l/\sigma_0$ play a role as the self-energy becomes insensitive to the wave number of the variation. In the limit of $\kappa R\to0$ the self-energy with a given $l$ goes to 0, the more so the larger the wave number, but with the scale still set by $\sigma_l$ and $\sigma_0$. Figure~\ref{fig:limits} shows the free energy correction of a charged shell due to local charge variation as given by Eq.~\eqref{eq:fcorr},
\begin{equation}
\label{eq:fcorr2}F_{DH}=F_{DH}^{(0)}(1+\gamma_{corr}),
\end{equation}
in the limits discussed above and for the special choice of $\widetilde{\sigma}_l/\sigma_0=1/|f_l^{min}|$. The local variation increases the shell self-energy, and quite significantly so in the limit $\kappa R\gg1$. On the other hand, the correction due to variation is smaller in the limit of $\kappa R\ll1$, and for the range of wave numbers relevant for the icosahedral symmetry is of the order of less than 1\% of the homogeneous contribution. The correction decreases with increasing wave numbers as smaller ``wavelengths'' of the extrema become smoothed out due to screening. The scattering of the data with respect to the wave numbers is a consequence of the minima of the functions $f_l^{min}$ changing non-monotonically with $l$.

\subsection{Interaction of two equal shells}

Let us finally turn attention to the interaction of two equal shells carrying charge distributions with icosahedral, octahedral, and tetrahedral symmetries. We will be mainly concerned with determining the minima of the interaction free energy as a function of the orientation of the shells. Two cases of inhomogeneous charge distributions will be explicitly considered: (i) the case of an overall neutral shell charge distribution and (ii) the same as (i) but with an added homogeneous background charge. Thus, when we will speak of a minimum energy it is assumed to be with respect to the shells' orientations unless stated otherwise.

Determining the orientation where the interaction energy has a minimum can be numerically demanding since the distributions have a varying number of local extrema the number of which increases with $l$, and that also become shallower at larger intershell separations. This in general increases the probability of finding only the local minimum of the interaction instead of the global one. However, we have implemented a minimization procedure based on simulated annealing~\cite{Frenkel,Corana1987} which reliably finds the global minimum in all of the cases we have considered, with an error in the interaction energy always below approximately 5\%.

\subsubsection{Neutral shells}

Considering the case of two neutral shells we have only one term in the interaction energy, $V_{ll}$, which depends on several parameters: the shells' orientations, their separation, as well as the wave number $l$, the corresponding symmetry, and the strength of the variation $\sigma_l$ (equal for both shells). Also variable are the bulk salt concentration $c_0$ and the radii of the shells $R$.

\begin{figure}[!htb]
\includegraphics[width=7.5cm]{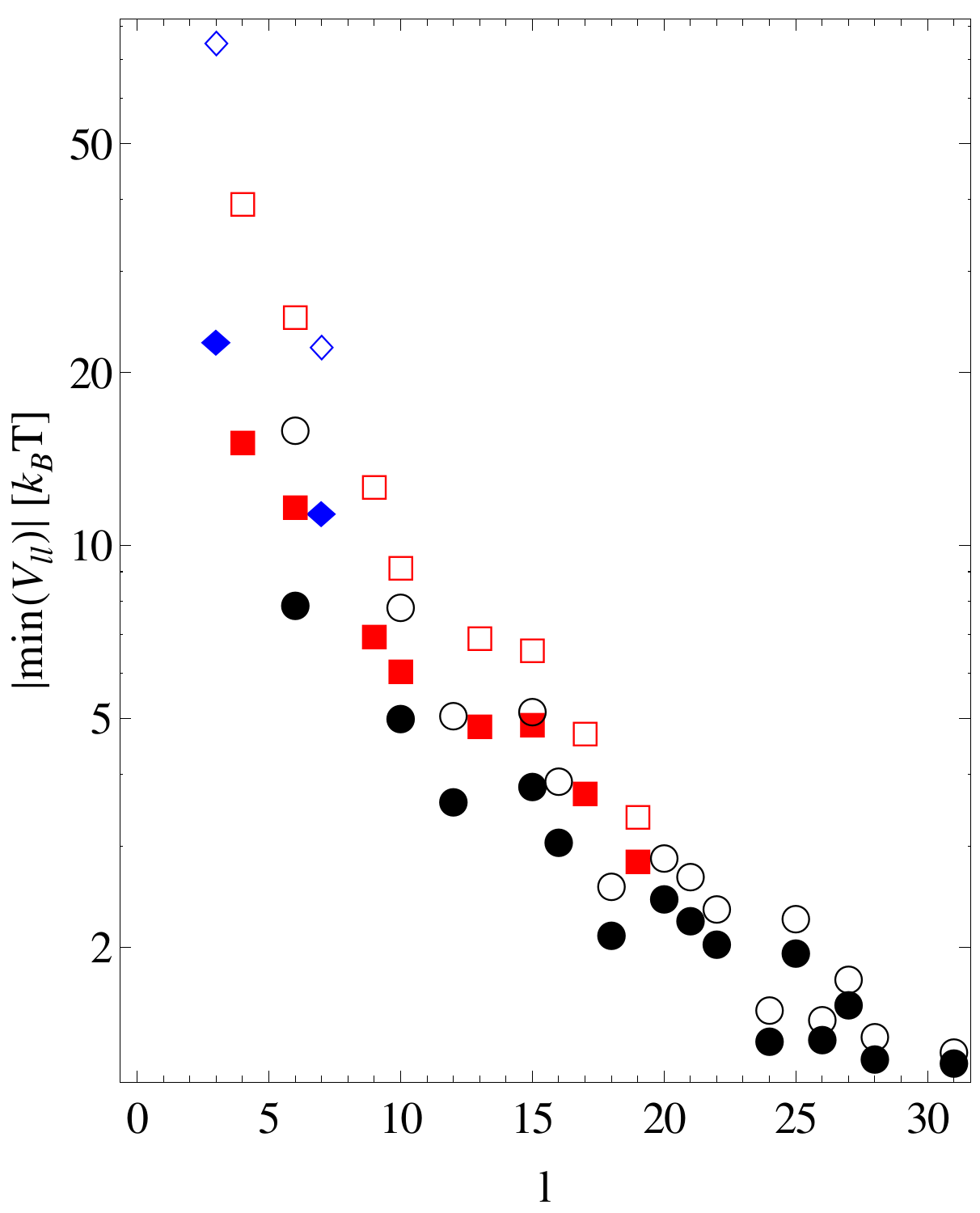}
\caption{Absolute value of the minimum interaction energy $V_{ll}$ at contact for two uncharged shells with symmetric variation (the interaction in the minimum is always attractive). Circles show wave numbers with icosahedral symmetry, squares ones with octahedral, and diamonds those with tetrahedral symmetry. Full symbols show the case where $\kappa R\approx10$ ($c_0=100$ mM and $R=10$ nm), and empty symbols the case with $\kappa R\approx3$ ($c_0=10$ mM and $R=10$ nm). Strength of the variation is taken to be $\sigma_l=1$ $e_0/\mathrm{nm}^2$.}
\label{fig:unchr_contact}
\end{figure}

To begin with, we show in Fig.~\ref{fig:unchr_contact} the absolute values of the minimum energy for the two shells at contact ($\rho=2R$); the interaction in the minimum is always attractive. The strength of the variation influences the scale of the energy in a predictable way, scaling with the square of $\sigma_l$ [Eq.~\eqref{eq:wlp}]. Lowering $\kappa R$ increases the interaction energy, as already observed in Sec.~\ref{sec:quad}, and the energy decreases with increasing $l$. Also, at a fixed $l$ the attraction in the minimum is stronger for a lower symmetry.

The $l$ dependence can be understood by considering the factors contributing to the interaction energy $V_{ll}$ in Eq.~\eqref{eq:wlp}, coming mainly from the function $C(l,l,\kappa R)$ and the modified spherical Bessel functions of the second kind $k_s(\kappa\rho)$, where the sum over $s$ is limited by $l$ (Appendix~\ref{app1:sec1}). The function $k_s$ takes on the biggest value when $s=l$, and thus the scaling of the energy is approximately set by the product of $C(l,l,\kappa R)$ and $k_l(\kappa\rho)$; the product which decreases as $l$ increases.

The symmetry dependence and to some extent the $l$ dependence as well are likely linked to the average size of the local charge patches with the same sign. Lower symmetries at a fixed $l$ have less minima, and the local patches of charge thus span a greater fraction of the shell surface. The interaction of two large patches brought close is less affected by neighboring patches some distance away, whereas the smaller patches of higher symmetries see an interaction averaged-out over several other patches in proximity.

\begin{figure}[!htb]
\includegraphics[width=7.5cm]{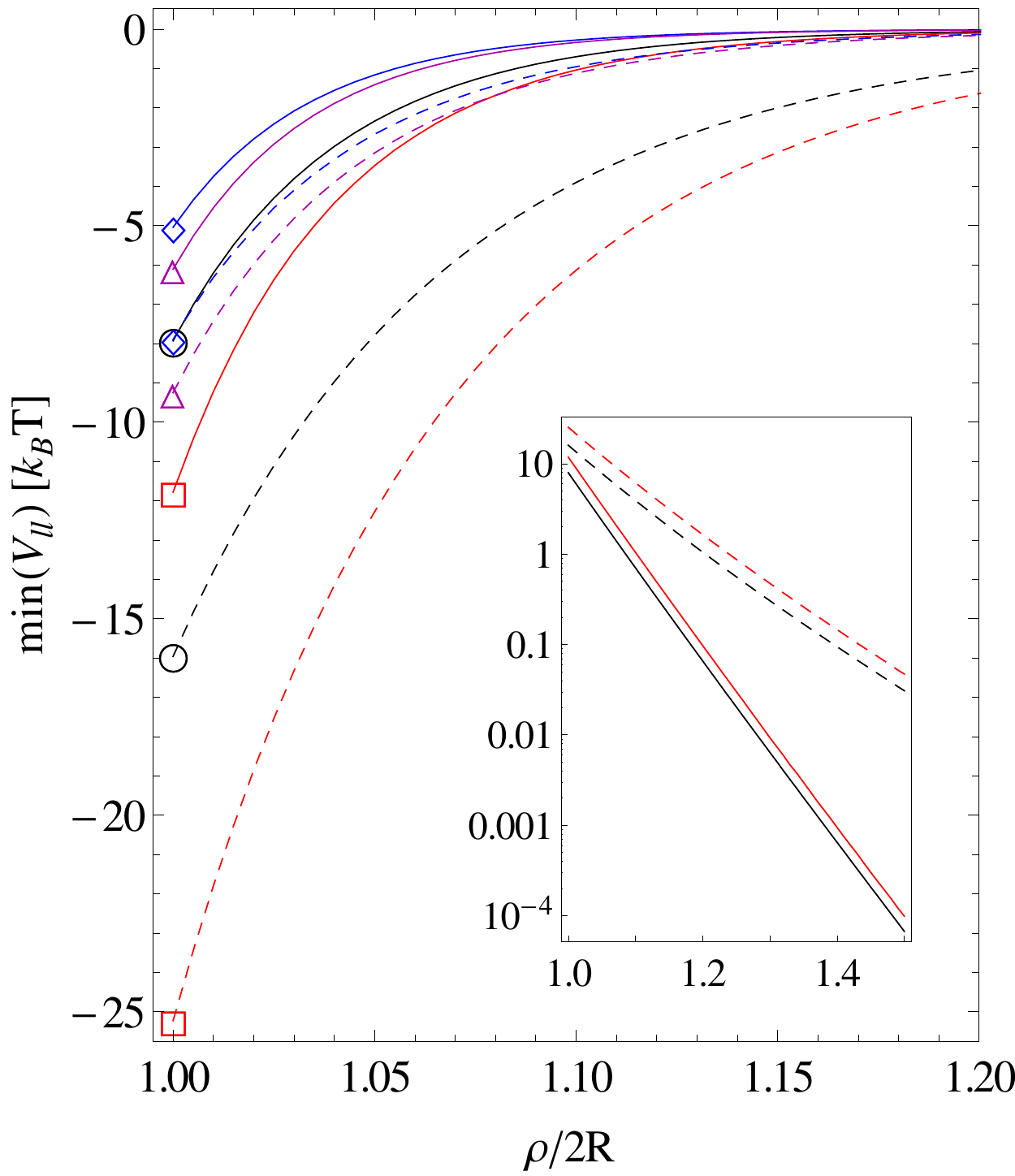}
\caption{Minimum of the interaction energy $V_{ll}$ as a function of the intershell separation for symmetries with even wave numbers: icosahedral symmetry with $l=6$ (circle) and $l=10$ (diamond) and octahedral symmetry with $l=6$ (square) and $l=10$ (triangle). Full lines show the case where $\kappa R\approx10$ and dashed lines a case with $\kappa R\approx3$. The inset shows the behavior of the absolute value of the energy for the two symmetries with $l=6$, showing the scaling at larger separations. Strength of the variation is taken to be $\sigma_l=1$ $e_0/\mathrm{nm}^2$.}
\label{fig:unchr_radial}
\end{figure}

\begin{figure}[!htb]
\includegraphics[width=7.5cm]{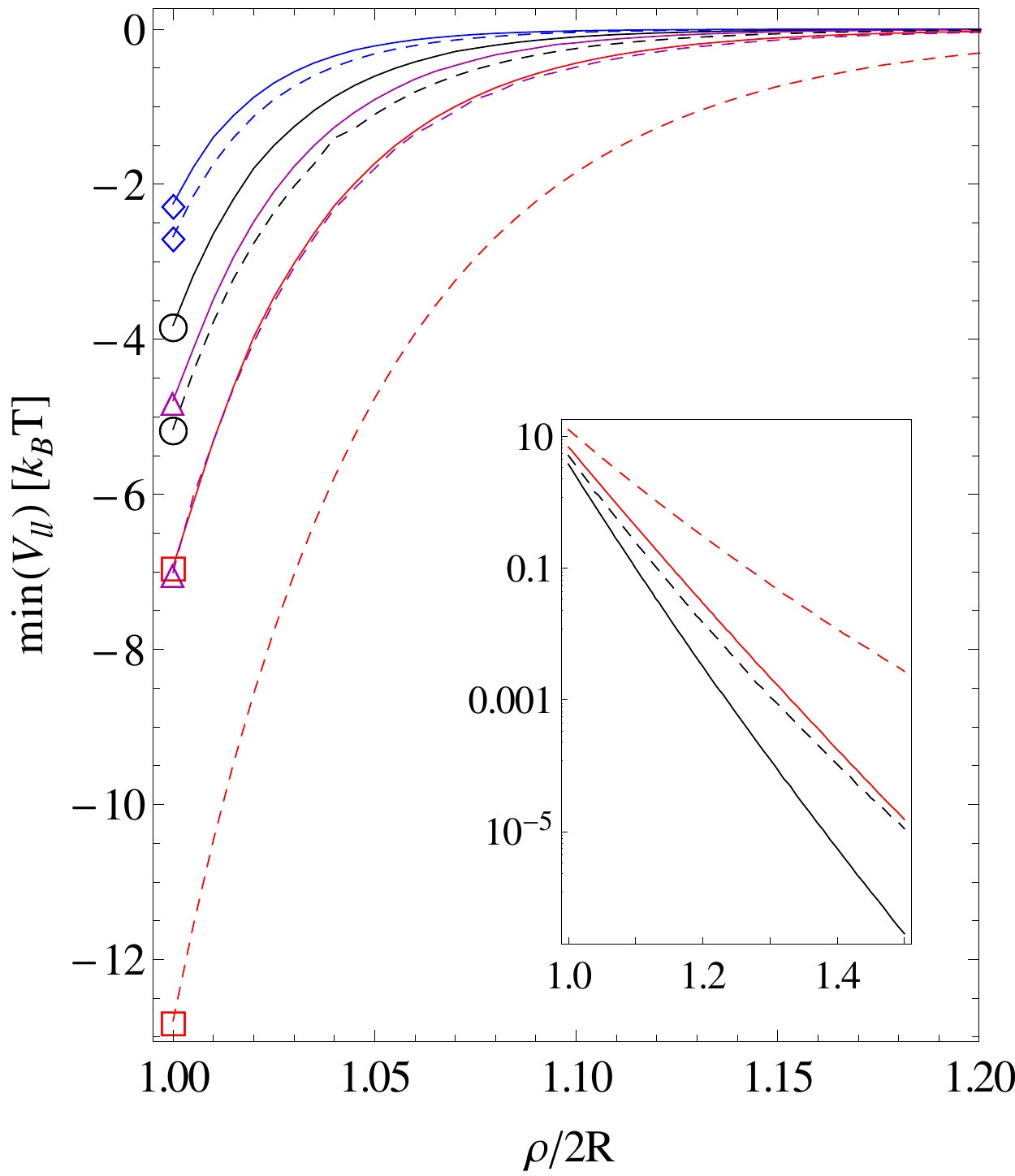}
\caption{Minimum of the interaction energy $V_{ll}$ as a function of the intershell separation for symmetries with odd wave numbers: icosahedral symmetry with $l=15$ (circle) and $l=21$ (diamond) and octahedral symmetry with $l=9$ (square) and $l=13$ (triangle). Full lines show the case where $\kappa R\approx10$ and dashed lines a case with $\kappa R\approx3$. The inset shows the behavior of the absolute value of the energy for $l=9$ and $l=15$, showing the scaling at larger separations. Strength of the variation is taken to be $\sigma_l=1$ $e_0/\mathrm{nm}^2$.}
\label{fig:unchr_radial2}
\end{figure}

Figures~\ref{fig:unchr_radial} and~\ref{fig:unchr_radial2} present the dependence of the interaction energy on the intershell separation for a number of different cases. The configuration of the two shells with the minimum energy does not appear to change with increasing separation; however, for every case there are several equal solutions due to the symmetry of the problem. We can observe again that lower $l$ and lower symmetries result in bigger attraction.

From Fig.~\ref{fig:unchr_radial} it is obvious that the scaling of the interaction energy with the intershell separation at a fixed wave number is influenced only by the dimensionless parameter $\kappa R$, with the energy falling off more rapidly with larger $\kappa R$. Changing the wave number also affects the scaling, as can be seen in Fig.~\ref{fig:unchr_radial2}, and the energy again falls off more rapidly with increased $l$.

Since the depths of minima and maxima of the surface charge distributions coincide for odd wave numbers and differ for even wave numbers, something similar is also observed in the interaction free energy when comparing different shell orientations. For odd $l$s the interaction energies in the minimum and maximum configuration mirror each other, even with changing separation, while for even $l$s the maximum of the interaction is usually somewhat larger than the minimum. For all the other orientations of the shells the energy falls in between the limits delineated by the minimal and maximal interaction, all the while keeping a monotonic behavior with respect to the intershell separation, i.e. being either purely repulsive or purely attractive.

\subsubsection{Charged shells}

On addition of a homogeneous background charge we can study the interaction of two charged shells in the presence of local charge density variation. The shells are assumed equal in size and the interaction energy is then
\begin{equation}
V_{int}(\rho,\boldsymbol{\omega}_i)=V_0(\rho)+W_{l0}(\rho,\boldsymbol{\omega}_i)+V_{ll}(\rho,\boldsymbol{\omega}_i),
\end{equation}
where $l$ is again chosen from the allowed wave numbers for a given symmetry. In addition to the parameters considered in the case of two neutral shells we can now modify the total charge on the shells by changing $\sigma_0$, and we can assume that the ratio $\sigma_l/\sigma_0$ will play an important role, similar to what we found out in Sec.~\ref{sec:quad}.

\begin{figure*}[!htb]
\includegraphics[width=17cm]{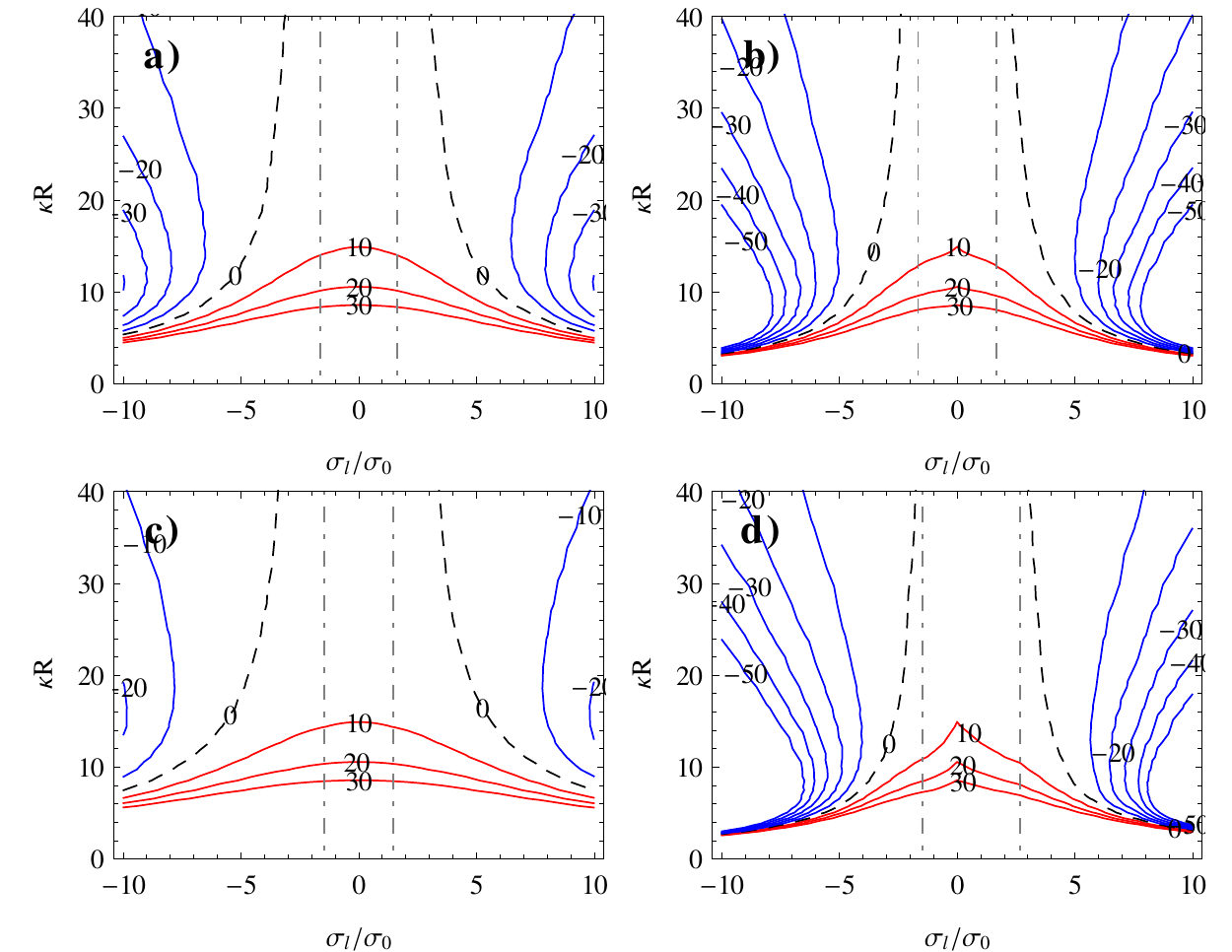}
\caption{Contour plots of the minimum of the interaction energy $V_{int}$ in units of $k_BT$ as a function of $\kappa R$ and the ratio $\sigma_l/\sigma_0$. The charge is kept fixed at $\sigma_0=0.4$ $e_0/\mathrm{nm}^2$, and we vary the variation $\sigma_l$. The shells are in contact, $\rho=2R$. Dashed lines show the contours where the interaction changes sign, and dot-dashed lines show the critical variation $\widetilde{\sigma}_l$ [Eq.~\eqref{eq:scrit}]. Below the critical variation the charge on the shells is everywhere of equal sign, irrespective of the variation. The wave numbers and symmetries are the same as in Fig.~\ref{fig:distributions}: {\bf a)} icosahedral with $l=15$, {\bf b)} octahedral with $l=9$, {\bf c)} icosahedral with $l=21$, and {\bf d)} dodecahedral with $l=6$.}
\label{fig:chr_contours}
\end{figure*}

Taking a look first at the two shells at contact ($\rho=2R$), we show in Fig.~\ref{fig:chr_contours} the contours of the minimum interaction energy as a function of $\kappa R$ and the ratio $\sigma_l/\sigma_0$ for the distributions already depicted in Fig.~\ref{fig:distributions}. Several things are notable: A sufficiently large charge variation with respect to $\sigma_0$ will cause a shift from repulsion to the attraction in the configuration of the two shells with the minimum energy. This critical ratio where the energy changes sign depends on $\kappa R$, and becomes increasingly large when $\kappa R$ approaches 0. On the other hand, for large $\kappa R$ the critical ratio seems to approach a constant, which is expectedly bigger than the critical ratio in Eq.~\eqref{eq:scrit} where the sign of the charge is still equal everywhere on the shells. Additionally, as already observed in previous cases, the sign of $\sigma_l$ has an effect only on distributions with even wave numbers, and the value of $\sigma_0$ simply sets the energy scale. However, the contours where the interaction energy changes sign remain the same regardless of the value of $\sigma_0$.

\begin{figure*}[!htb]
\includegraphics[width=17cm]{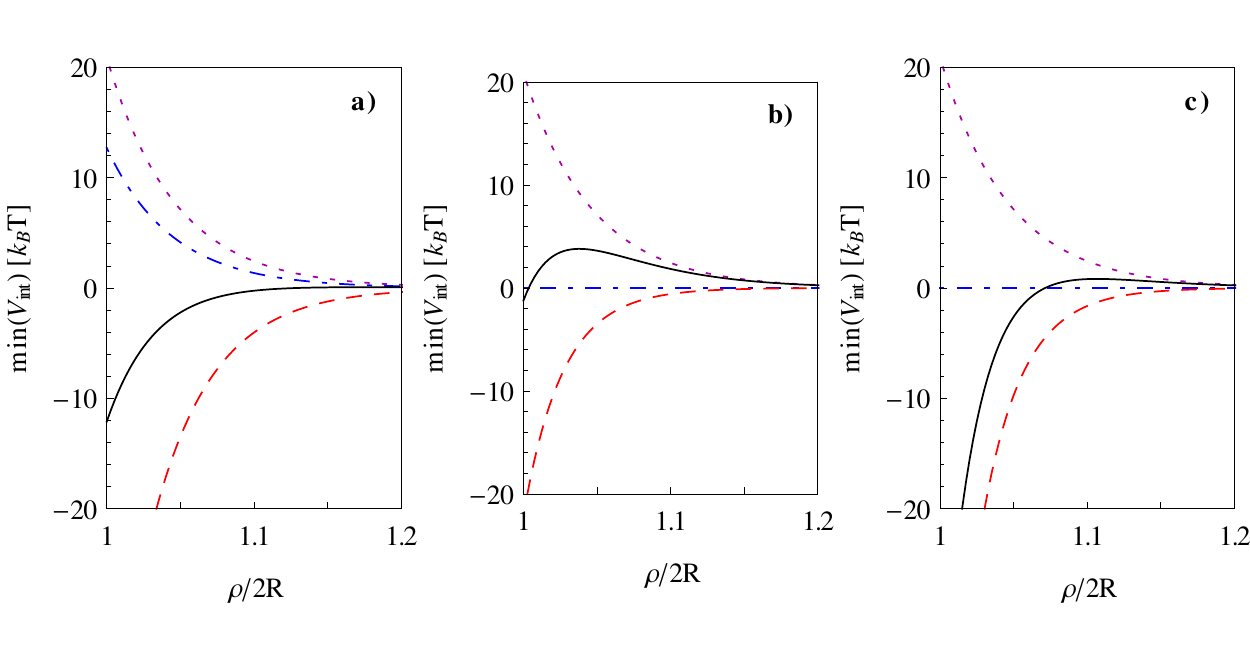}
\caption{Total interaction energy $V_{int}$ of two charged shells with icosahedral variation (full line) and the relative contributions ($V_0$ -- dotted line, $W_{l0}$ -- dot-dashed line, $V_{ll}$ -- dashed line) as a function of the intershell separation for {\bf a)} $l=6$ and $\sigma_l/\sigma_0=6$, {\bf b)} $l=15$ and $\sigma_l/\sigma_0=6$, and {\bf c)} $l=15$ and $\sigma_l/\sigma_0=10$. The radii of the shells are $R=10$ nm with $\kappa R\approx10$, and the charge on the shells is fixed to $\sigma_0=0.4$ $e_0/\mathrm{nm}^2$, making the $V_0$ contribution equal for all three cases.}
\label{fig:ico_radial}
\end{figure*}

The major influence on the critical ratio where the minimum of the interaction energy changes sign appears again to be the number of extrema in the distribution, where lower number of extrema needs lesser variation strength to achieve attraction -- meaning lower $l$ and lower symmetries. In distributions with the same symmetry and similar $l$ the minima of those with odd wave numbers are slightly stronger compared to distributions with even $l$. This can be understood by taking a look at the cross-term $W_{l0}$, where one can observe from Eq.~\eqref{eq:wlp} that a factor of $(-1)^l$ appears in the summation over the expansion coefficients. Thus, in the minimum configurations of distributions with odd $l$ the cross-term is several orders of magnitude lower than the $V_{0}$ and $V_{ll}$ contributions, whereas the cross-term is comparable if the $l$ is even (Fig.~\ref{fig:ico_radial}).

From this Section we can conclude that even like-charged particles can be found in configurations where they attract each other, if only the local variation of charge is large enough. It is the presence of patches of opposite sign that mediates this attraction, as can be seen from the regions excluded by dot-dashed lines in Fig.~\ref{fig:chr_contours}. The attraction sets in for close separations of the two particles, as the long-range behavior is dominated by the repulsion of the homogeneous contribution, however small (Fig.~\ref{fig:ico_radial}). Any configuration other than the one with minimum energy can show either a purely repulsive or a mixed short-range attractive/long-range repulsive behavior, as already observed in Sec.~\ref{sec:quad}.

\section{Discussion and Conclusions}\label{sec:con}

We have derived an analytical expression for the interaction of two permeable, arbitrarily charged spherical particles in the DH approximation. To study the interactions of virus-like particles or other shells with symmetric variations in surface charge distributions we have introduced model charge distributions with icosahedral, octahedral, and tetrahedral symmetry, and characterized by a single wave number.

We find that local variations in the charge distributions leading to local charge separation into oppositely charged patches reduce the repulsion between like-charged shells, a feature that has also been observed in models of protein-protein interactions~\cite{Grant2001} as well as in DNA-DNA interactions~\cite{Kornyshev1997,Korny-rev}. An important consequence of the reduced repulsion is that it can be furthermore turned into explicit attraction if the background homogeneous charge is small enough. This onset of attraction between like-charged particles is due to the presence of local variation of charge, since the attracting patches have to have opposite sign. The attraction is short-ranged with respect to the separation between the particles, with the long-ranged interaction engendered by the homogeneous charge background still being (vanishingly) repulsive. Apart from the minimum interaction free energy configuration, any other configuration can exhibit either a purely repulsive or a mixed short-range attractive/long-range repulsive behavior.

One of the drawbacks of the approach used here is the linearization of the non-linear Poisson-Boltzmann (PB) equation. The approximation is well-justified for radii relevant for viruses~\cite{ALB2012} ($R\gtrsim10$ nm) and salt concentrations of $c_0\sim100$ mM. However, for smaller salt concentrations where $\kappa R\sim1$ the DH theory overestimates the values of the potential and consequently the energy~\cite{Siber2007,ALB2011}, meaning that the values obtained for the interaction energy at shell contact are too big. The qualitative behavior, on the other hand, is expected to be much the same~\cite{Siber2007}.

The non-homogeneous charge distributions introduced in this paper were all mapped on a single-shell capsid model. A further improvement would be a two-shell model, with each shell carrying a distribution corresponding to the charge on the inner and outer shells~\cite{ALB2012}. Nonetheless, the single-shell model is able to capture the main aspects of the electrostatic interactions in the system~\cite{Siber2007}.

If the charge distributions can be characterized by one or only a few different wave numbers (like the ones considered in this paper), this greatly simplifies the expressions for the interaction potential, limiting many of the sums involved. On the other hand, distributions where patches extend over well-defined surface regions can be difficult to be parametrized within the formalism. The most extreme cases would be distributions of point charges on a shell or a partially formed capsid as considered in Ref.~\onlinecite{ALB2011} -- i.e. distributions consisting of delta functions or Heaviside step functions in the solid angle, which are also notoriously difficult to accurately represent with a finite number of terms in an expansion.

The symmetric charge distributions presented can serve as improved models in a broad variety of virus systems. One such example could be the adsorption of flexible polyelectrolytes on viral capsids which have dodecahedral charge distribution, where a combination of the lowest dodecahedral symmetry ($l=6$ in Fig.~\ref{fig:distributions}) with the addition of a homogeneous background nicely fits the model of discrete charges as used in Ref.~\onlinecite{Linse2007}. These models also open up the possibilities of studying the interaction of viral capsids with multivalent ions within the dressed counterions theory, or even multivalent ion mediated interaction between two such capsids~\cite{Kanduc2010a,Kanduc2011,Javidpour??}.

The derived expressions for the interaction potential are certainly too heavy to be used in computer simulations. And even though molecular dynamics simulations show that switching from continuous to discrete charge patterns in model spherical proteins leads to significant differences between the resulting screened Coulomb interactions~\cite{Hoffmann2004}, the continuous distributions presented are more tractable analytically, especially since the symmetries can be described by using one or at most few different wave numbers.

\begin{acknowledgments}
A.L.B. is grateful to Luka Leskovec for helpful discussions and Antonio \v{S}iber for comments on the original manuscript.

A.L.B. acknowledges the financial support from the Slovene Agency for Research and Development under the young researcher grant. R.P. acknowledges the financial support from the Slovene Agency for Research and Development under grants P1-0055 and J1-4297.
\end{acknowledgments}

\appendix

\section{Details of the derivation}\label{app:app1}

\subsection{Properties of Wigner 3-j symbols}\label{app1:sec1}

The combination of the two Wigner 3-j symbols in the function ${\cal L}_{lm}^{pq}$ [Eq.~\eqref{eq:llm}] gives the rise to following properties~\cite{Rose1957,Gray}: the function is non-zero only when $m=q$ and $l+p+s=\mathrm{even}$. The product of the two Wigner 3-j symbols also makes the function ${\cal L}_{lm}^{pq}$ invariant under all odd and even permutations of the three wave numbers $l$, $s$, and $p$. The sum over $s$ is limited with the triangle inequality, $|l-s|\leq p\leq l+s$. This further restricts $s$ to even numbers when $l=p$; on the other hand, when one of them is zero (e.g. $p=0$), the sum contains only one term, $s=l$.

\subsection{Obtaining coefficients for the electrostatic potential of the shells}\label{app1:sec2}

Writing out the continuity equation [Eq.~\eqref{eq:bc1}] on the surface of the second shell in terms of the functions $\psi_{lm}^\pm$ we obtain
\begin{eqnarray}
\nonumber&&\sum_{l,m}a(lm,2)i_l(\kappa R)Y_{lm}(\Omega)=\\
\nonumber&&=\sum_{l,m}b(lm,2)k_l(\kappa R)Y_{lm}(\Omega)+\\
&&+\sum_{l,m}b(lm,1)\sum_{p,q} {\cal L}_{lm}^{pq}(1)i_p(\kappa R)Y_{pq}(\Omega).
\end{eqnarray}
By multiplying both sides with $Y_{l'm'}^*$ and  using the orthogonality properties of spherical harmonics~\cite{Abramowitz1965}, we have after integrating over $\Omega$
\begin{equation}
\label{eq:bc1a}
\left[a(lm,2)-\sum_{p,q}b(pq,1){\cal L}_{pq}^{lm}(1)\right]i_l(\kappa R)=b(lm,2)k_l(\kappa R).
\end{equation}
With the same procedure the boundary condition for the discontinuity of the electric field [Eq.~\eqref{eq:bc2}] on the surface of the second shell on the other hand yields
\begin{eqnarray}
\label{eq:bc2a}
&&\nonumber\left[a(lm,2)-\sum_{p,q}b(pq,1){\cal L}_{pq}^{lm}(1)\right]i_l'(\kappa R)-\\
&&-b(lm,2)k_l'(\kappa R)=\frac{\sigma(lm,2)}{\kappa\varepsilon\varepsilon_0}.
\end{eqnarray}
The factors in the square braces of Eqs.~\eqref{eq:bc1a} and~\eqref{eq:bc2a} are the same, giving
\begin{equation}
b(lm,2)\frac{k_l(\kappa R)}{i_l(\kappa R)}i_l'(\kappa R)-b(lm,2)k_l'(\kappa R)=\frac{\sigma(lm,2)}{\kappa\varepsilon\varepsilon_0},
\end{equation}
from which we obtain the coefficients $b(lm,2)$ of the exterior solution [Eq.~\eqref{eq:blm2}]. Same approach is used to obtain $b(lm,1)$. By inserting these coefficients back into the boundary condition equations, we then get the coefficients of the interior solution, Eq.~\eqref{eq:alm2}.

\begin{widetext}
\subsection{Deriving the symmetrized terms of interaction energy}\label{app1:sec3}

The function $W_{lp}$ we have introduced in Eq.~\eqref{eq:wlp} written out in full is
\begin{eqnarray}
\nonumber W_{lp}&=&\frac{R^2}{2\kappa\varepsilon\varepsilon_0}\sum_{m,q}\Bigg\{\frac{C_0(p,\kappa R)i_l(\kappa R)}{k_p(\kappa R)}\Big[\sigma(pq,1)\sigma^*(lm,2){\cal L}_{pq}^{lm}(1)+\sigma^*(lm,1)\sigma(pq,2){\cal L}_{pq}^{lm}(2)\Big]+\\
&+&\frac{C_0(l,\kappa R)i_p(\kappa R)}{k_l(\kappa R)}\Big[\sigma(lm,1)\sigma^*(pq,2){\cal L}_{lm}^{pq}(1)+\sigma^*(pq,1)\sigma(lm,2){\cal L}_{lm}^{pq}(2)\Big]\Bigg\}.
\end{eqnarray}
We can separate the four terms of the sum into two parts, $W_{lp}=W_{lp}^{(1)}+W_{lp}^{(2)}$, where
\begin{equation}
W_{lp}^{(1)}=\frac{R^2}{2\kappa\varepsilon\varepsilon_0}\sum_{m,q}\left[\frac{C_0(p,\kappa R)i_l(\kappa R)}{k_p(\kappa R)}\sigma(pq,1)\sigma^*(lm,2){\cal L}_{pq}^{lm}(1)+\frac{C_0(l,\kappa R)i_p(\kappa R)}{k_l(\kappa R)}\sigma^*(pq,1)\sigma(lm,2){\cal L}_{lm}^{pq}(2)\right]
\end{equation}
and similarly for $W_{lp}^{(2)}$. The Wigner 3-j symbols appearing in the functions ${\cal L}_{lm}^{pq}$ are non-zero only if $q=m$. Since the functions are invariant under odd and even permutations we get that ${\cal L}_{pm}^{lm}(1)$ and ${\cal L}_{lm}^{pm}(2)$ differ only in the factor $(-1)^s$ appearing in the sum in the latter case. Thus the sum over $q$ disappears and using the fact that $l+p+s=\mathrm{even}$, we get
\begin{eqnarray}
\nonumber W_{lp}^{(1)}&=&\frac{R^2}{2\kappa\varepsilon\varepsilon_0}\sum_{m,s}(-1)^{p+m}(2s+1)\sqrt{(2l+1)(2p+1)}\,k_s(\kappa\rho)\left(\begin{array}{ccc}
      l & s & p \\
      0 & 0 & 0
      \end{array}
\right)
\left(\begin{array}{ccc}
      l & s & p \\
      \underline{m} & 0 & m
      \end{array}
\right)\times\\
&\times&\Bigg[\frac{C_0(p,\kappa R)i_l(\kappa R)}{k_p(\kappa R)}\sigma(pm,1)\sigma^*(lm,2)+\frac{C_0(l,\kappa R)i_p(\kappa R)}{k_l(\kappa R)}\sigma^*(pm,1)\sigma(lm,2)\Bigg].
\end{eqnarray}
Additionally we can use the following relations:
\begin{eqnarray}
\frac{C_0(p,\kappa R)i_l(\kappa R)}{k_p(\kappa R)}+\frac{C_0(l,\kappa R)i_p(\kappa R)}{k_l(\kappa R)}&=&2C(l,p,\kappa R),\\
\frac{C_0(p,\kappa R)i_l(\kappa R)}{k_p(\kappa R)}-\frac{C_0(l,\kappa R)i_p(\kappa R)}{k_l(\kappa R)}&=&0,
\end{eqnarray}
where the function $C(l,p,x)$ is defined in Eq.~\eqref{eq:aux}, and after some manipulation we obtain
\begin{eqnarray}
\nonumber W_{lp}^{(1)}&=&\frac{R^2}{\kappa\varepsilon\varepsilon_0}C(l,p,\kappa R)\sum_{m,s}(-1)^{p+m}\Re[\sigma(pm,1)\sigma^*(lm,2)]\times\\
&\times&(2s+1)\sqrt{(2l+1)(2p+1)}\,k_s(\kappa\rho)
\left(\begin{array}{ccc}
      l & s & p \\
      0 & 0 & 0
      \end{array}
\right)
\left(\begin{array}{ccc}
      l & s & p \\
      \underline{m} & 0 & m
      \end{array}
\right).
\end{eqnarray}
In an analogous manner we can derive a similar expression for $W_{lp}^{(2)}$, and putting both results together we retrieve Eq.~\eqref{eq:wlp}.
\end{widetext}

\section{Limiting cases of interaction energy}\label{app:app2}

In this Appendix we shall derive some limiting cases of the general expression for the interaction energy to show that we obtain the results found in the literature. A useful partial result is the expression for the interaction of two homogeneously charged shells, $V_{0}$.

The distribution is spherically symmetric and invariant under all rotations, thus giving for both shells $\sigma(00,i)=\sqrt{4\pi}\sigma_0$ (this normalization of the coefficient is due to the requirement that $4\pi R^2\sigma_0=Q$ gives the total charge on a shell). Taking this into account, the interaction of two such shells is found to be
\begin{equation}
\label{eq:V00}V_0(\rho)=\frac{4\pi R^2\sigma_0^2}{\kappa\varepsilon\varepsilon_0}\sinh^2(\kappa R)\times\frac{e^{-\kappa\rho}}{\kappa \rho}.
\end{equation}

\subsection{Coulomb limit}

Firstly, we take a look at the Coulomb limit where $\kappa R\to 0$. In this limit, the function $C(l,p,x)$ found in partial terms of the interaction energy $W_{lp}$ [Eq.~\eqref{eq:wlp}] goes as
\begin{equation}
\lim_{x\ll 1}C(l,p,x)=x^{l+p}\frac{2^{-(l+p+1)}x^2}{\Gamma(l+3/2)\Gamma(p+3/2)},
\end{equation}
where $\Gamma(z)$ is the Gamma function~\cite{Abramowitz1965}. The bigger the wave number of the charge variation, the smaller role it plays in this limit, and in the lowest order of the expansion in $x=\kappa R$ we need focus only on the lowest wave numbers $l=p=0$. Thus we obtain using Eq.~\eqref{eq:V00}
\begin{equation}
\lim_{\kappa R\to0}V(\rho,\boldsymbol{\omega}_i)\approx\lim_{\kappa R\to0}V_{0}(\rho)=\frac{Q^2}{4\pi\varepsilon\varepsilon_0}\times\frac{e^{-\kappa\rho}}{\rho},
\end{equation}
the screened Coulomb (DH) interaction of two point charges $Q$.

\subsection{Limit of large screening}

Another limit is the regime of large screening, $\kappa R\to\infty$. We need to consider the asymptotic expansions of the function $C(l,p,x)$ and the modified spherical Bessel functions appearing in the expression for the interaction energy:
\begin{eqnarray}
\lim_{x\gg 1}C(l,p,x)&=&\frac{e^{2x}}{2\pi}+\frac{e^{-2x}(-1)^{l+p}}{2\pi}-\frac{(-1)^l+(-1)^p}{2\pi}.\\
\label{eq:appks}k_s(\kappa\rho)&\asymp&\frac{\pi}{2\kappa}\frac{e^{-\kappa H}}{H+2R}e^{-2\kappa R},
\end{eqnarray}
where we have written the distance between the shell centers $\rho$ in terms of their distance of closest contact $H$, $\rho=H+2R$. From Eq.~\eqref{eq:appks} it is obvious that in this limit the dominant term in the radial dependence is independent of the wave number, but the anisotropy of the two distributions persists~\cite{Hoffmann2004}. This makes it hard to obtain any general results as the interaction in general depends on the orientations of the two shells when they are brought together. But if we consider two shells at small separations, $\rho\ll R$, we can check the interaction of two homogeneously charged shells [Eq.~\eqref{eq:V00}] and obtain
\begin{equation}
V_0(H)=\frac{R\sigma_0^2}{8\kappa^2\varepsilon\varepsilon_0}e^{-\kappa H},
\end{equation}
which has the same functional dependence as given by~\citet{Verwey}, but differs in the prefactor due to a different approximation.

\subsection{Dipole-dipole interaction}\label{app2:sec3}

Another limit we can check is the expression for the interaction of two (equal) dipole distributions. We concern ourselves with an axially symmetric charge distribution with $l=1$, $\sigma(1m,i)=\sigma_1\delta_{m0}$. The connecting line between the two dipoles is chosen to be $\hat{\mathbf{z}}$, and due to the symmetry of the problem the only relevant Euler angles are the azimuthal angles of the two dipoles, $\beta_1$ and $\beta_2$. The dipole moment of the shells can be calculated from their surface charge distributions~\cite{Gray},
\begin{equation}
\boldsymbol{\mu}=\int\mathbf{r}\,\sigma(\Omega)\,\delta(r-R)\,\mathrm{d}^3\mathbf{r}.
\end{equation}
The distribution in the reference frame is axially symmetric, and we get a dipole moment in the $z$ direction only with the magnitude
\begin{equation}
\mu=\sigma_1R^3\sqrt{\frac{4\pi}{3}}.
\end{equation}
Thus, we can write for the general orientations of the two dipoles $\boldsymbol{\mu}_1\boldsymbol{\mu}_2=\mu^2(\cos\beta_1\cos\beta_2+\sin\beta_1\sin\beta_2)$ and $(\boldsymbol{\mu}_1\hat{\mathbf{z}})(\boldsymbol{\mu}_2\hat{\mathbf{z}})=\mu^2\cos\beta_1\cos\beta_2$. After summing over $m$ and $s$ the dipole-dipole interaction energy is
\begin{eqnarray}
\nonumber V_{11}&=&\frac{R^2\sigma_1^2}{\kappa\varepsilon\varepsilon_0}C(1,1,\kappa R)\left(-\frac{3\pi e^{-\kappa\rho}}{2(\kappa\rho)^3}\right)\times\\
\nonumber&\times&\Big[(2+2\kappa\rho+(\kappa\rho)^2)\cos\beta_1\cos\beta_2-\\
&&-(1+\kappa\rho)\sin\beta_1\sin\beta_2\Big].
\end{eqnarray}
Simplifying the above equation using the expressions for the dipole moment we can write in the Coulomb limit of two point dipoles
\begin{eqnarray}
\nonumber\lim_{\kappa R\to 0}V_{11}/k_BT&=&l_B\frac{e^{-\kappa\rho}}{\rho^3}\Big[(1+\kappa\rho)\boldsymbol{\mu}_1\boldsymbol{\mu}_2-\\
&-&(3+3\kappa\rho+(\kappa\rho)^2)(\boldsymbol{\mu}_1\hat{\mathbf{z}})(\boldsymbol{\mu}_2\hat{\mathbf{z}})\Big],
\end{eqnarray}
obtaining the same expression for the DH interaction of two point dipoles as given in the literature (e.g. Ref.~\onlinecite{Kumar2009}).

\bibliographystyle{apsrev.bst}

\end{document}